\documentclass[12pt,preprint]{aastex}

\newcommand{\oi}{[O\,{\footnotesize I}]}
\newcommand{\oiii}{[O\,{\footnotesize III}]}
\newcommand{\cii}{[C\,{\footnotesize II}]}
\newcommand{\nii}{[N\,{\footnotesize II}]}
\newcommand{\siii}{[Si\,{\footnotesize II}]}
\newcommand{\suiii}{[S\,{\footnotesize III}]}
\newcommand{\suiv}{[S\,{\footnotesize IV}]}
\newcommand{\feii}{[Fe\,{\footnotesize II}]}
\newcommand{\feiii}{[Fe\,{\footnotesize III}]}
\newcommand{\ariii}{[Ar\,{\footnotesize III}]}
\newcommand{\neii}{[Ne\,{\footnotesize II}]}
\newcommand{\neiii}{[Ne\,{\footnotesize III}]}
\newcommand{\piii}{[P\,{\footnotesize III}]}
\newcommand{\hii}{H\,{\footnotesize II}}
\newcommand{\um}{\,$\mu$m}
\newcommand{\cc}{\,cm$^{-3}$}

\shorttitle{Si and Fe depletion in Galactic star-forming regions}
\shortauthors{Okada et al.}

\begin{document}


\title{Si and Fe depletion in Galactic star-forming regions \\
    observed by the {\it Spitzer Space Telescope}}


\author{Yoko Okada\altaffilmark{1}, Takashi Onaka\altaffilmark{2}, Takashi Miyata\altaffilmark{3}, Yoshiko K. Okamoto\altaffilmark{4}, Itsuki Sakon\altaffilmark{2}, Hiroshi Shibai\altaffilmark{5} and Hidenori Takahashi\altaffilmark{6}}


\altaffiltext{1}{Institute of Space and Astronautical Science, Japan Aerospace Exploration Agency, 3-1-1 Yoshinodai, Kanagawa 229-8520, Japan; okada@ir.isas.jaxa.jp}
\altaffiltext{2}{Department of Astronomy, Graduate School of Science, University of Tokyo, 7-3-1 Hongo, Bunkyo-ku, Tokyo 113-0033, Japan}
\altaffiltext{3}{Institute of Astronomy, Graduate School of Science, University of Tokyo, 2-21-1 Osawa, Mitaka, Tokyo 181-0015, Japan}
\altaffiltext{4}{Institute of Astrophysics and Planetary Science, Faculty of Science, Ibaraki University, 2-1-1, Bunkyo, Mito, Ibaraki 310-8512, Japan}
\altaffiltext{5}{Division of Particle and Astrophysical Science, Graduate School of Science, Nagoya University, Furo-cho, Chikusa-ku, Nagoya 464-8602, Japan}
\altaffiltext{6}{Gunma Astronomical Observatory, 6860-86 Nakayama, Takayama, Agatsuma, Gunma 377-0702, Japan}


\begin{abstract}
We report the results of the mid-infrared spectroscopy of 14 Galactic star-forming regions with the high-resolution modules of the Infrared Spectrograph (IRS) on board the {\it Spitzer Space Telescope}.  We detected \siii\ 35\um, \feii\ 26\um, and \feiii\ 23\um\ as well as \suiii\ 33\um\ and H$_2$ S(0) 28\um\ emission lines.  Using the intensity of \nii\ 122\um\ or 205\um\ and \oi\ 146\um\ or 63\um\ reported by previous observations in four regions, we derived the ionic abundance Si$^+$/N$^+$ and Fe$^+$/N$^+$ in the ionized gas and Si$^+$/O$^0$ and Fe$^+$/O$^0$ in the photodissociation gas.  For all the targets, we derived the ionic abundance of Si$^+$/S$^{2+}$ and Fe$^{2+}$/S$^{2+}$ for the ionized gas.  Based on photodissociation and \hii\ region models the gas-phase Si and Fe abundance are suggested to be 3--100\% and $<8$\% of the solar abundance, respectively, for the ionized gas and 16--100\% and 2--22\% of the solar abundance, respectively, for the photodissociation region gas.  Since the \feii\ 26\um\ and \feiii\ 23\um\ emissions are weak, the high sensitivity of the IRS enables to derive the gas-phase Fe abundance widely in star-forming regions.  The derived gas-phase Si abundance is much larger than that in cool interstellar clouds and that of Fe.  The present study indicates that 3--100\% of Si atoms and $<22$\% of Fe atoms are included in dust grains which are destroyed easily in \hii\ regions, probably by the UV radiation.  We discuss possible mechanisms to account for the observed trend; mantles which are photodesorbed by UV photons, organometallic complexes, or small grains.
\end{abstract}



\section{Introduction}

One of the methods to probe chemical compositions of interstellar dust grains is to examine the gas-phase abundance of a certain element and attribute its deficiency against the reference abundance (i.e. depletion) to those contained in dust grains.  Since the gas-phase emission or absorption lines are relatively easy to observe and interpret, this method can be applied for large samples.  The depletion pattern in diffuse interstellar matter (ISM) has been studied on the basis of ultraviolet (UV) absorption lines \citep{SavageSembach96,Jenkins04}.  Previous work for tenuous interstellar clouds ($n_\mathrm{H} \lesssim 10$\cc) indicates that Si, Mg, and Fe, which are the major constituents of dust grains, show different depletion patterns; Si atoms return to the gas phase most easily, Mg is depleted more than or similarly to Si, and Fe atoms tend to remain most insistently in dust grains \citep{Sofia94, Fitzpatrick96, Jones00, Cartledge04, Cartledge06, Miller07}.  These suggest that silicates contain primarily Mg and most or all Fe atoms are in other components of dust such as metal or oxides.  The depletion of these elements correlates with the average hydrogen density in the line of sight, indicating that the ISM consists of two types of clouds, warm and cold \citep{Cartledge06,Jensen07}.  For more dense clouds with $n_\mathrm{H}=100$--1000\cc, Si and Fe is highly depleted \citep{Joseph86}.

Depletion in active star-forming regions, where the density is much higher than in diffuse clouds ($n_\mathrm{H} \sim 10$--$10^6$\cc), has been studied by infrared line emissions with the {\it Kuiper Airborne Observatory} ({\it KAO}) and the {\it Infrared Space Observatory} \citep[{\it ISO};][]{Kessler96}.   Intense \siii\ 35\um\ emission has been reported in a shocked region in Orion \citep{Haas91}, the Galactic center region \citep{Stolovy95}, and the massive star-forming region Sharpless 171 \citep[S171,][]{Okada03} and the Carina nebula \citep{Mizutani04}, which are attributed to the large gas-phase Si abundance of $>30$\% of solar.  Not only in ionized regions but also in photodissociation regions (PDRs) large gas phase Si abundances of 10--50\% of solar have been reported in the massive star-forming region S171,  G333.6-0.2 \citep{Colgan93} and OMC-1 \citep{Rosenthal00}, and in reflection nebula $\rho$ Oph \citep{Okada06} and NGC~7023 \citep{Fuente00}, whereas \citet{YoungOwl02} indicated that the intensities of \siii\ 35\um\ in two reflection nebulae agree with the interstellar gas phase abundance (a few percent of solar).

At visual wavelengths, abundance studies of \hii\ regions have been carried out for the Galactic gradients of O \citep{Shaver83,Pilyugin03}, and also of N, S, or C \citep{Vilchez96,Esteban05} in the context of the Galactic chemical evolution.  Several observations of Chephides and OB stars have also been performed to examine the abundance in the Galactic disk.  The derived O abundance in the solar vicinity is compatible with its gas-phase abundance in \hii\ regions \citep[][ and references therein]{Rolleston00,Pilyugin03}.  There are few optical studies of the gas-phase abundance of highly depleted elements in \hii\ regions.  \citet{Rodriguez02} shows that Fe abundance in seven Galactic star-forming regions is estimated to be 2--30\% of solar.  Since the \feii\ 26\um\ and \feiii\ 23\um\ lines are weak, only a few detections of these lines have been so far reported \citep{Timmermann96,vandenAncker00,Rosenthal00} and little constraint on the gaseous Fe abundance has been imposed before the {\it Spitzer Space Telescope} \citep{Werner04} in infrared.  In this paper, we report the depletion pattern of Fe as well as Si in Galactic star-forming regions observed with the Infrared Spectrograph \citep[IRS;][]{Houck04} on board {\it Spitzer}.

\section{Observations and Data Reduction}

\subsection{Target Selection}

The properties of the targets and the observation parameters are listed in Table~\ref{targetsummary} and the coordinates of each observed position are listed in Table~\ref{lineLH}.  In the following, s\# represents the observed position identifiers.  We have observed two groups of targets: (a) \hii~region/PDR complexes within a few kpc from the Sun that have been studied with {\it ISO} or {\it KAO} mapping observations, and (b) giant \hii~regions in the Galactic plane located at various distances from the Galactic center and on different spiral arms.  As group (a), we have selected four star-forming regions where far-infrared spectroscopic observations have been made by {\it ISO} or {\it KAO} at more than two positions, and the \nii~122\um\ or 205\um\ (if in the ionized region) and \oi~146\um~or 63\um~emissions have been obtained, and the properties of the PDR gas have been derived.  Among about twenty regions that satisfy these criteria, we have selected four regions that covers a wide range of the energy of the exciting source (Table~\ref{targetsummary}).  We examine correlations among lines detected by the IRS with them, deriving the abundance of Si$^+$ and Fe$^+$ relative to N$^+$ or O$^0$ in \hii~regions and/or PDRs.  To take account of the difference in the aperture size between the IRS ($11.1\arcsec\times 22.3\arcsec$ for the whole slit) and the {\it ISO}/LWS ($66.4\arcsec$--$87.0\arcsec$) or {\it KAO} ($45\arcsec$--$60\arcsec$), we observed 4 positions ($2\times 2$ mapping) with the IRS within the aperture of the {\it ISO} or {\it KAO}.  

Group (b) targets have been selected from \citet{Conti04} of a catalog of giant Galactic \hii\ regions with a Lyman continuum photon number $N(\mathrm{Lyc})\geq 10^{50}$ photon s$^{-1}$.  Bright and giant regions are suitable for mapping observations within a reasonable observation time.  To minimize the uncertainty from the density variation in the observed regions, we have selected \hii\ regions that show smooth radio continuum emission and do not have clumpy structures in them.  

\subsection{Observed positions in individual targets}

S171 is an active star-forming region excited by an O7 type star \citep[see ][and the references therein]{Okada03}.  \citet{Okada03} have made one-dimensional raster scan observations of p1--p24, going from the vicinity of the exciting star (p1) to the molecular clouds (p24).  We observed 6 positions, each of which has $2\times 2$ mapping, of p8--p10 and p19--p21 in \citet{Okada03}, where p8 corresponds to s0--s3 of the present study, p9 to s4--s7, and so on (Table~\ref{lineLH}).  The ionized gas is dominant at p8--p10, where \nii\ 122\um\ emission has a local maximum and located at the edge of a molecular cloud \citep{Okada03}.  Contribution from the PDR gas is suggested to be significant at p19--p21, around the edge of another molecular cloud.  

$\sigma$ Sco is a star-forming region excited by a B1 type star, which is not associated with CO molecular clouds \citep[see ][and the references therein]{Okada06}.  \citet{Okada06} have made one-dimensional raster scan observations of p1--p15, through $\sigma$ Sco (between p13 and p14) toward the north-west direction, where the \cii\ 158\um\ and {\it IRAS} 100\um\ continuum emissions show an extended structure.  We observed p7, p9, and p10 in \citet{Okada06}, where \nii\ 122\um\ emission has been detected, as s0--s3, s4--s7, and s8--s11, respectively, in the present study (Table~\ref{lineLH}).   

\begin{figure}
\epsscale{1.0}
\plotone{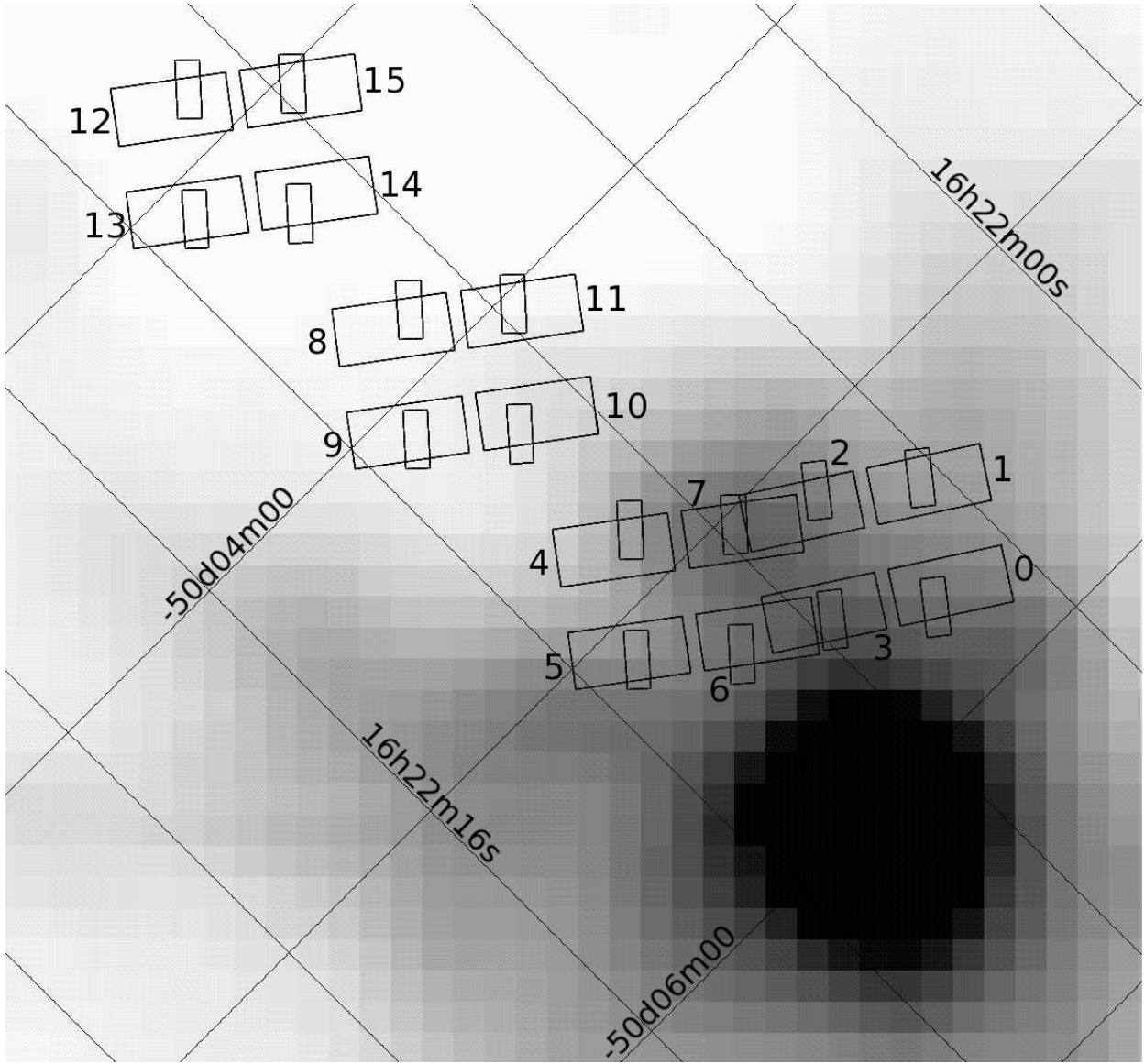}
\caption{The observed positions for G333.6-0.2 on the  {\it MSX} 8\um\ with reversed gray-scale.  The larger rectangles indicate the slit of the LH module and the smaller ones that of the SH module.  The numbers labeled to each slit position correspond to s\# in Tables~\ref{lineLH} and \ref{lineSH}.\label{G333obspos}}
\end{figure}

G333.6-0.2 is a bright \hii~region at radio and infrared wavelengths and almost totally obscured in the visible.  It is likely to be excited by a cluster of O and B stars \citep{Goss70,Becklin73,Fujiyoshi98}.  Since the center of G333.6-0.2 is too bright to observe, we made observations at 3 positions (s4--s7, s8--s11, and s12--s15) that correspond to N1, N2, and N3 in \citet{Simpson04}, $1\arcmin$, $2\arcmin$, and $3\arcmin$ away from the center to the north (Fig.~\ref{G333obspos} and Table~\ref{lineLH}).  The positions of s0--s3 are located at $\sim 1\arcmin$ northwest from the center.  Part of the Long-High (LH) field-of-view of s2 overlaps with that of s7, and so does s3 with s6 (Fig.~\ref{G333obspos}).  

NGC~1977 is excited by a B1 star and has an \hii\ region/molecular cloud interface that is nearly edge-on \citep[][and the references therein]{Subrahmanyan01,YoungOwl02}.  We observed at the \cii\ peak position, the \oi\ peak position, and the position with the second strongest \oi\ intensity in Table 4 of \citet{YoungOwl02} as s0--s3, s4--s7, and s8--s11, respectively (Table~\ref{lineLH}).  Among them the \cii\ peak is the nearest ($\sim 6\arcmin$ southwest) from the exciting star.  All those positions are located at outside of the optically bright edge, which corresponds to the ionization front \citep{Kutner79,Subrahmanyan01}.

For group (b) targets, we made staring observations with the nodding at 3 or 4 positions for each target.  The positions of s0 and s1 are a nodding pair, so are s2 and s3, s4 and s5, and s6 and s7.  For two targets, {\it ISO}/SWS and LWS single position observations have been reported by \citet{Peeters02} (see Table~\ref{lineLH}).

\subsection{Observation parameters and techniques}

The observations were performed in the cycle 2 general observer program (ID: 20612) of {\it Spitzer}.  All targets were observed with the LH module of the IRS, which provided spectra of $\lambda/\Delta\lambda\sim 600$ in 18.7--37.2\um.  Some of the targets were observed also with the Short-High (SH) module in 9.9--19.6\um, which are listed in Table~\ref{lineSH}.  For S171 (except for s4--s7), NGC~1977, G32.797+0.192, and G351.467-0.462, we made background observations to reduce the effects of rogue pixels\footnote{See IRS Data Handbook}.  Other targets are bright enough and we do not need background observations, or the observations had been made before the problem of rogue pixels was noticed.  For these targets the background subtraction is not applied.

\begin{figure}
\epsscale{1.0}
\plotone{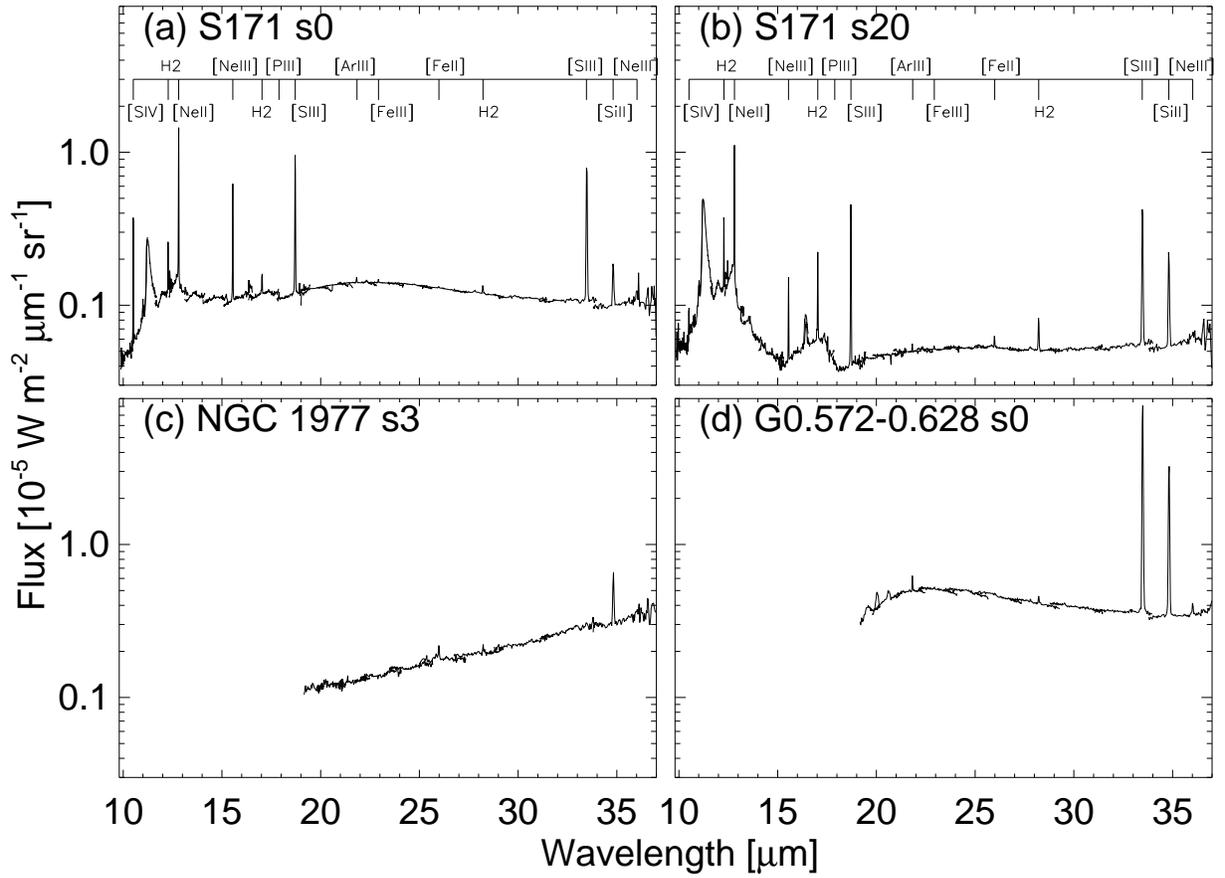}
\caption{Typical examples of the spectra of \hii/PDR complexes, a) s0 of S171, b) s20 of S171, c) s3 of NGC~1977, and d) s0 of G0.572-0.628.\label{spectra}}
\end{figure}

For all the observations, the data of the pipeline version S13 or S14, which are identical for LH and SH, were used.  We start with the basic calibrated data (BCD).  When we made background observations, we subtract them from the on-source observations.  After we apply IRSCLEAN to correct rogue pixels, we process the BCD by using SMART\footnote{SMART was developed by the IRS Team at Cornell University and is available through the Spitzer Science Center at Caltech.} software \citep{Higdon04}.  After coadding the BCD of the same positions, we extract the spectra with the full aperture, i.e. taking a sum along the slit with SMART.  Since the IRS intensities have been calibrated against point sources, we apply the extended source correction, which takes account of the aperture correction and the slit efficiency.  Examples of the obtained spectra are shown in Fig.~\ref{spectra}.

We obtained the line intensities by Gaussian fits by assuming a linear baseline.  The results are summarized in Tables~\ref{lineLH} and \ref{lineSH}.  The errors in the line intensities include both the fitting and the baseline uncertainties.  For the line emission less than 3 times the uncertainty ($<3\sigma$) we give $3\sigma$ as an upper limit.  For each mapping dataset of group (a), we coadded the spectra and derive average line intensities to compare with the results of {\it ISO} and {\it KAO}.  They are also shown in these tables.  In the following, we describe these coadded spectra as s(0--3), to be distinguished from individual observations from s0 to s3 that are designated by s0--s3.

\subsection{Origin of individual emission lines}

Table~\ref{IPtable} shows the ionization potential of elements of the detected lines and the configuration and the energy level of upper state are given in Table~\ref{lineproptable}.  Among emission lines detected by the present observations, \siii\ 35\um\ and \feii\ 26\um\ can originate both from the ionized and PDR gas since the first ionization potentials of Si and Fe are less than 13.6~eV.  On the other hand, \feiii\ 23\um\ as well as \suiii\ 33\um\ and \nii\ 122\um\ stem only from the ionized gas thus we use those lines to discuss the gas-phase abundance in the ionized gas.  These lines are collisionally excited by electrons.  The H$_2$ S(0) 28\um\ emission as well as \oi\ 63\um\ and 146\um\ traces the PDR gas.

\subsection{Extinction corrections and foreground contamination}

We apply extinction correction to the observed line intensities as follows.  S171, $\sigma$ Sco, and NGC~1977 are located within 1~kpc from the Sun and well visible in optical \citep{Sharpless59,Whittet74,Subrahmanyan01}.  The mean $A_V$ per kiloparsecs for lines of sight close to the Galactic plane is $\sim 1.8$ mag \citep{Whittet03}, and thus the extinction is considered to be negligible at mid-infrared wavelengths.  For G333.6-0.2, we estimate the extinction correction factor to be multiplied to the observed intensities from $\tau_{9.7}=1.5$ \citep{Fujiyoshi01,Simpson04} and the extinction law from \citet{WD01} and \citet{Draine03a,Draine03b} to be 3.4 for \suiv~10.5\um\ and 1.3 for \siii~35\um.  Group (b) targets are all located in the inner Galaxy and $>5$~kpc from the Sun, and thus large extinction is expected.  However, we mainly discuss only the line ratios within the LH wavelength range.  Even with $A_\mathrm{V}=60$ mag, which is about two times of the average visual extinction towards the Galactic center, the \siii~35\um/\suiii~33\um\ ratio and the \feiii\ 23\um/\suiii\ 33\um\ ratio becomes larger only by -6\% and 50\%, respectively.  Therefore, we do not apply the extinction corrections to the group (b) targets and take account of the uncertainty of $-6$\% and $+50$\% for the \siii~35\um/\suiii~33\um\ and the \feiii\ 23\um/\suiii\ 33\um\ ratio, respectively.

For group (b) targets, the obtained line intensities can be contaminated by foreground gas in the Galactic plane.  We estimate the effect from the 1420~MHz continuum observations of NRAO VLA sky survey \citep{Condon98}.  The intensity of the radio continuum of the position observed by the IRS is larger than the surrounding diffuse emission by a factor of 10 or more, except for G32.797+0.192 s4 and s5 and G359.429-0.090 s4--s7.  For G32.797+0.192 s4 and s5 the factor is $\sim 8$, while the radio intensity at G359.429-0.090 s4--s7 is almost the same as diffuse emissions.  Therefore, we consider the obtained line intensities at G359.429-0.090 s4--s7 as upper limits and exclude them in the following abundance discussion.

\section{Results}

\begin{figure}
\epsscale{0.45}
\plotone{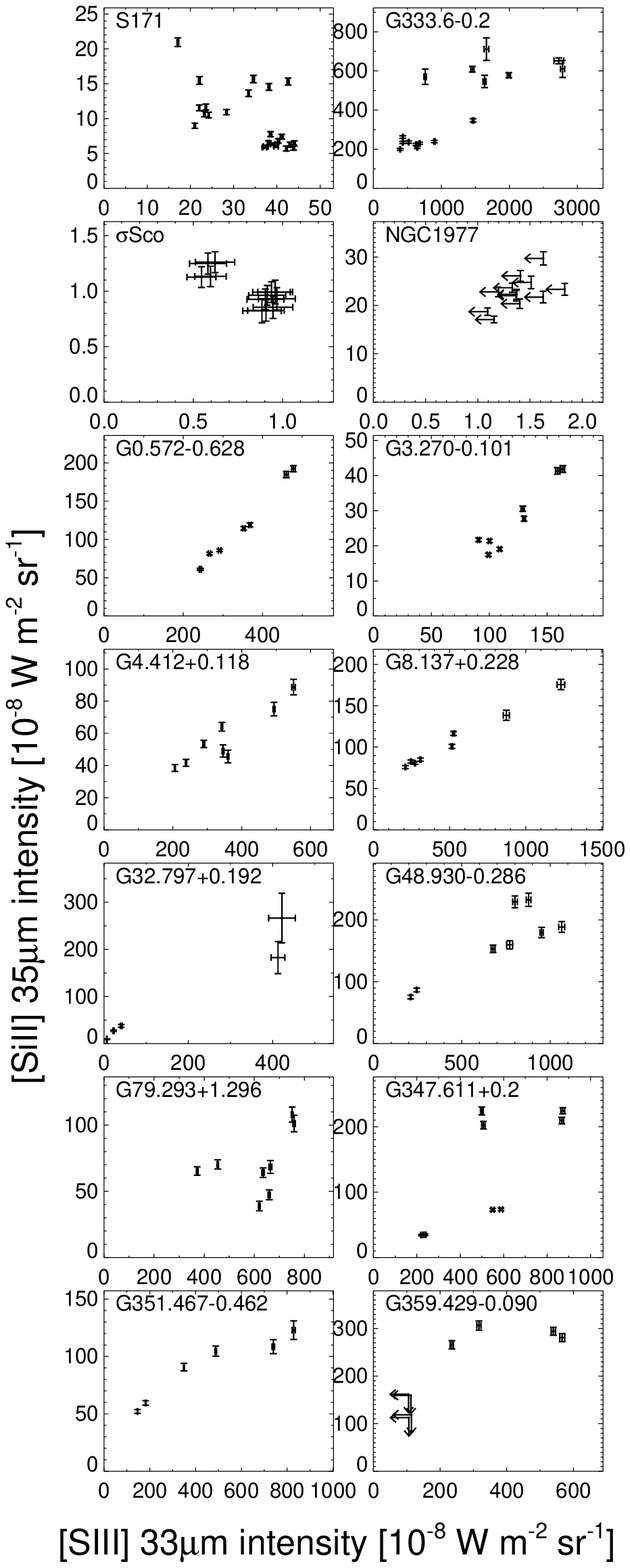}
\plotone{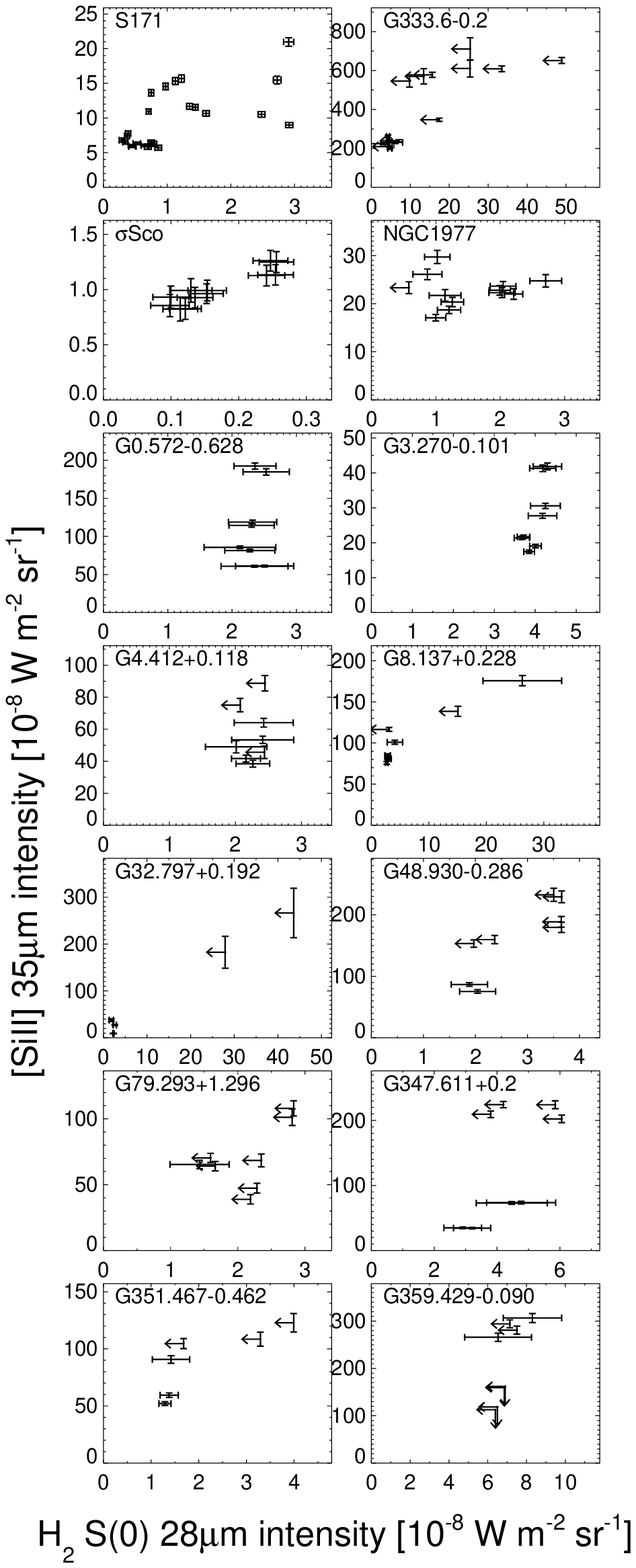}
\caption{\siii\ 35\um\ intensity versus \suiii\ 33\um\ intensity (left) and H$_2$ S(0) 28\um\ intensity (right) for each targets.\label{SIIIH2SiII35}}
\end{figure}

\begin{figure}
\epsscale{0.45}
\plotone{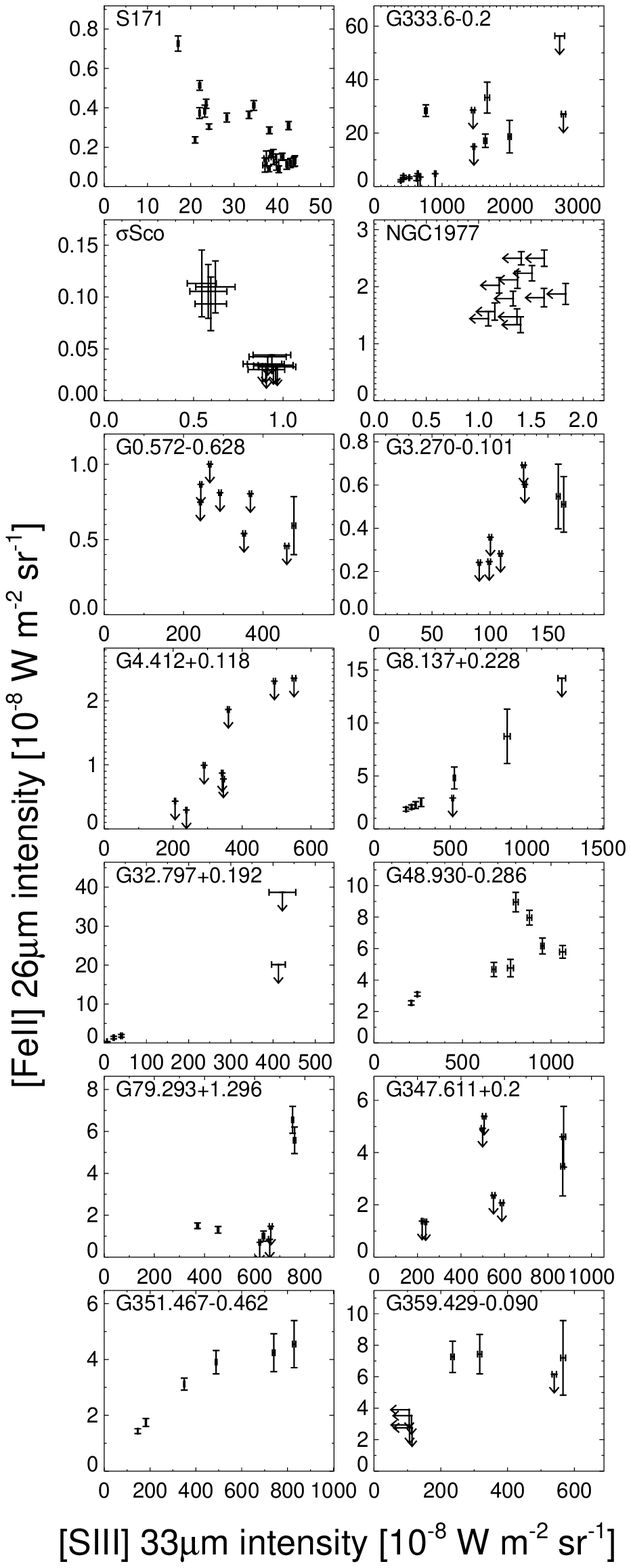}
\plotone{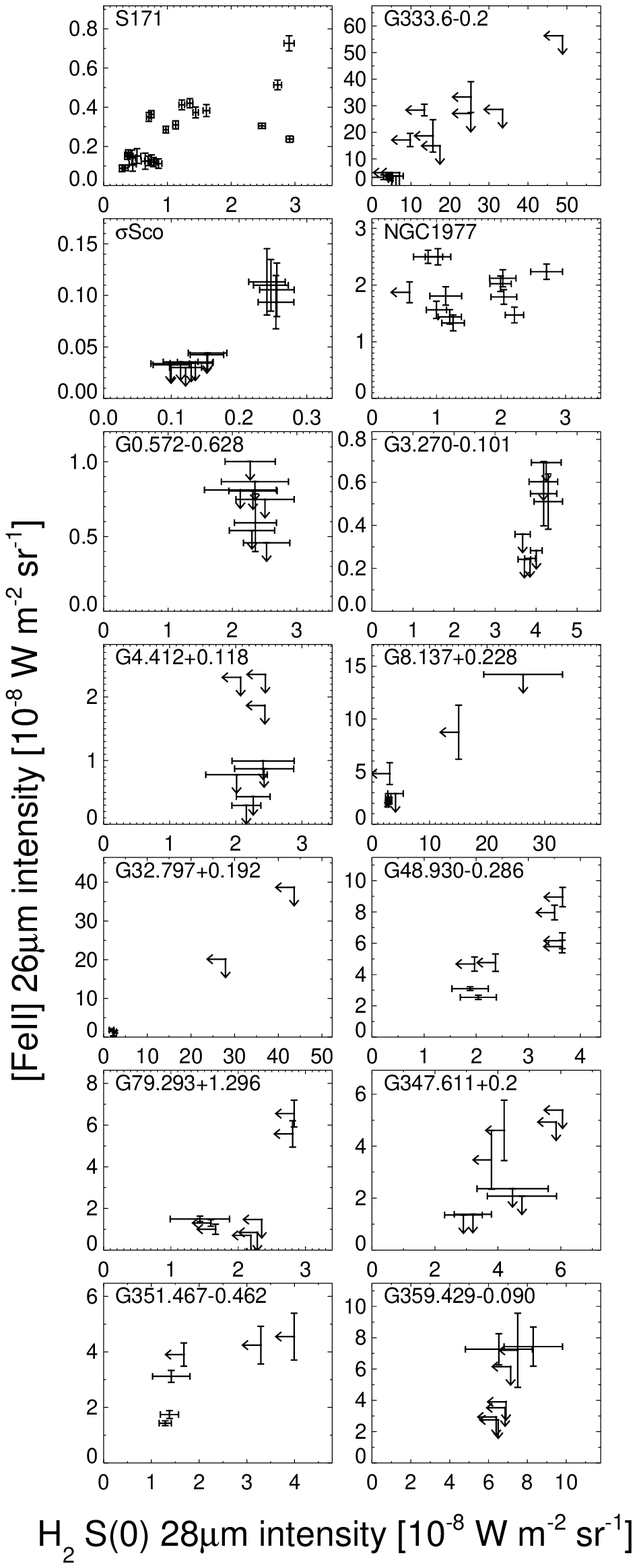}
\caption{\feii\ 26\um\ intensity versus \suiii\ 33\um\ intensity (left) and H$_2$ S(0) 28\um\ intensity (right) for each targets.\label{SIIIH2FeII26}}
\end{figure}

The results are summarized in Tables~\ref{lineLH} and \ref{lineSH}.  \siii\ 35\um\ emission is detected at all the positions, and \feii\ 26\um\ and \feiii\ 23\um\ emissions are detected at more than half of the mapping positions of most targets.  We examine the correlation of \siii\ 35\um\ and \feii\ 26\um\ against \suiii\ 33\um\ or H$_2$ S(0) 28\um, which traces the ionized gas and PDR gas, respectively (Figs.~\ref{SIIIH2SiII35} and \ref{SIIIH2FeII26}), to investigate the origin of the \siii\ 35\um\ and \feii\ 26\um\ emission.  The correlation coefficients are estimated for data where both lines are detected, and listed in Table~\ref{corr_coeff}.  Since the number of data is small, we make t-test to examine the reliability of the derived correlation coefficients, and the results are also shown in Table~\ref{corr_coeff}.  When the rejection region is $<1$\%, the assumption that there is no correlation can be rejected with 99\% reliability.

In the following, we estimate the abundance ratio between several ions using these line ratios, taking account of the contributions from the ionized and PDR gas.  For the ionized gas, we refer to models of an isothermal cloud of 7000--12000~K with a uniform density with the optically thin condition.  The relevant fine structure levels are taken into account except for Fe$^+$, for which a6D, a4F, and a4D levels are also included into the calculation.  Pumping and subsequent downward cascades from higher levels are not taken into account, but they have negligible effects on the \feii\ 26\um\ emission \citep{Verner00}.  Collision coefficients and A-coefficients of relevant transitions are from \citet{Zhang95} and \citet{Quinet96} for Fe$^+$, \citet{Nahar96} and \citet{Zhang96} for Fe$^{2+}$, \citet{Osterblock89} and \citet{Shibai92} for Si$^+$, N$^+$, and S$^{2+}$.  Using the density of each region derived by previous studies (Table~\ref{targetsummary}), we estimate the abundance ratio of two ions from the line ratio.  In the following, we describe the abundance ratio of two ions against the solar abundance by \citet{Asplund05} shown in Table~\ref{abundance_table} as (Fe$^{2+}$/S$^{2+}$)$_\mathrm{as}\equiv$ (Fe$^{2+}$/S$^{2+}$)/(Fe$^{2+}$/S$^{2+}$)$_\odot$, where ``as'' means ``against solar''.  Since S and N are considered to be little depleted \citep{SavageSembach96}, we use line ratios against \suiii\ or \nii\ lines to discuss the depletion pattern of Si and Fe.  For the PDR gas, we compare the observed line ratios with those of PDR models by \citet{Kaufman06} and simply attribute the discrepancy to the abundance ratios.  The assumed abundance of each element in the PDR model is indicated in Table~\ref{abundance_table}.  We estimate the abundance ratio from comparison of the observed line ratios with those of the models by simple scaling.

\subsection{Group (a)}

\subsubsection{S171}

The \siii\ 35\um\ intensities derived from the averaged spectra over 4 mapping positions (Table~\ref{lineLH}) agree with that of {\it ISO}/SWS \citep{Okada03} within $1\sigma$ except for s(12--15) and s(20--23), where they are in agreement within $2\sigma$.  The difference in the aperture size is the most likely cause of the disagreement, since the variation within each 4 mapping positions is larger than the uncertainty of the line intensity derived from the averaged spectrum.  The \siii\ 35\um\ and \nii\ 122\um\ emissions observed with {\it ISO} show a good correlation, suggesting that most \siii\ 35\um\ emission comes from the ionized gas except for p19 and p21, where a significant contribution from the PDR is indicated \citep{Okada03}.  Figure~\ref{SIIIH2SiII35} and Table~\ref{corr_coeff} show that the \siii\ 35\um\ emission correlates with the H$_2$ 28\um\ and it shows rather a negative correlation with the \suiii\ 33\um\ emission.  The positions of the IRS observations are limited around a local maximum of \siii\ 35\um\ and \nii\ 122\um\ and do not cover a wide range along the distance from the exciting star.  The positions of s0--s11 (p8--p10) is located at 1.4--1.8~pc from the exciting star and s12--s23 (p19--p21) is located at 3.6--4.0~pc.  Both correspond to clouds around dense molecular clouds, and the contribution from the PDR gas is suggested for the latter from the {\it ISO} observations.  Thus the present correlation is severely affected by local conditions and does not directly imply the origin of the line emission.  \citet{Okada03} suggested that (Si$^+$/N$^+$)$_\mathrm{as}=0.3\pm 0.1$ assuming that the ionized gas for the origin of the \siii\ 35\um, which is (Si$^+$/N$^+$)$_\mathrm{as}=0.22\pm 0.07$ when we use the solar abundance in Table~\ref{abundance_table}.

\begin{figure}
\epsscale{1.0}
\plotone{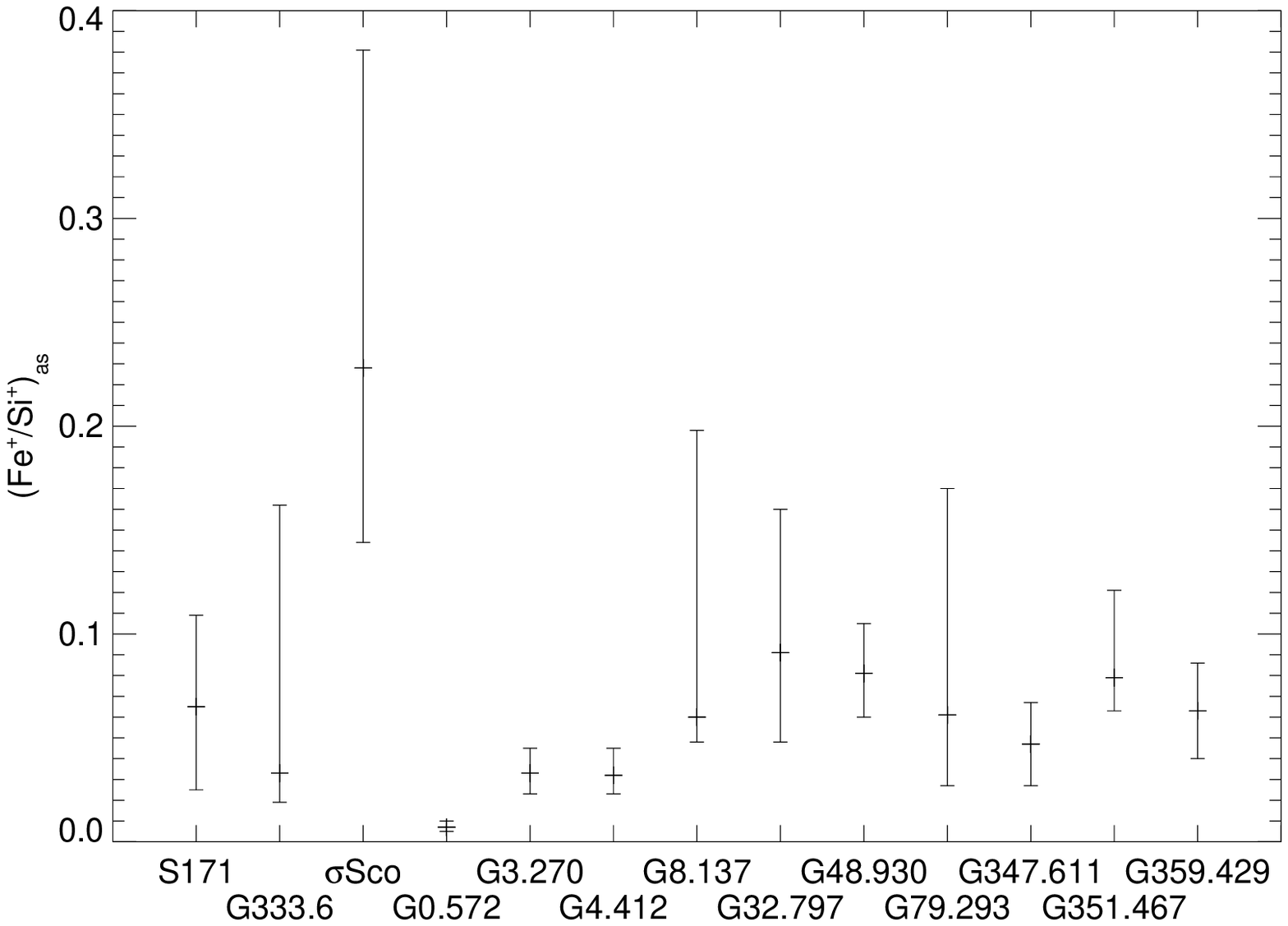}
\caption{(Fe$^{+}$/Si$^{+}$)$_\mathrm{as}$ with assuming that \feii\ 26\um\ and \siii\ 35\um\ emissions are from the ionized gas.\label{FeII26SiII35_figure}}
\end{figure}

\begin{figure}
\epsscale{1.0}
\plotone{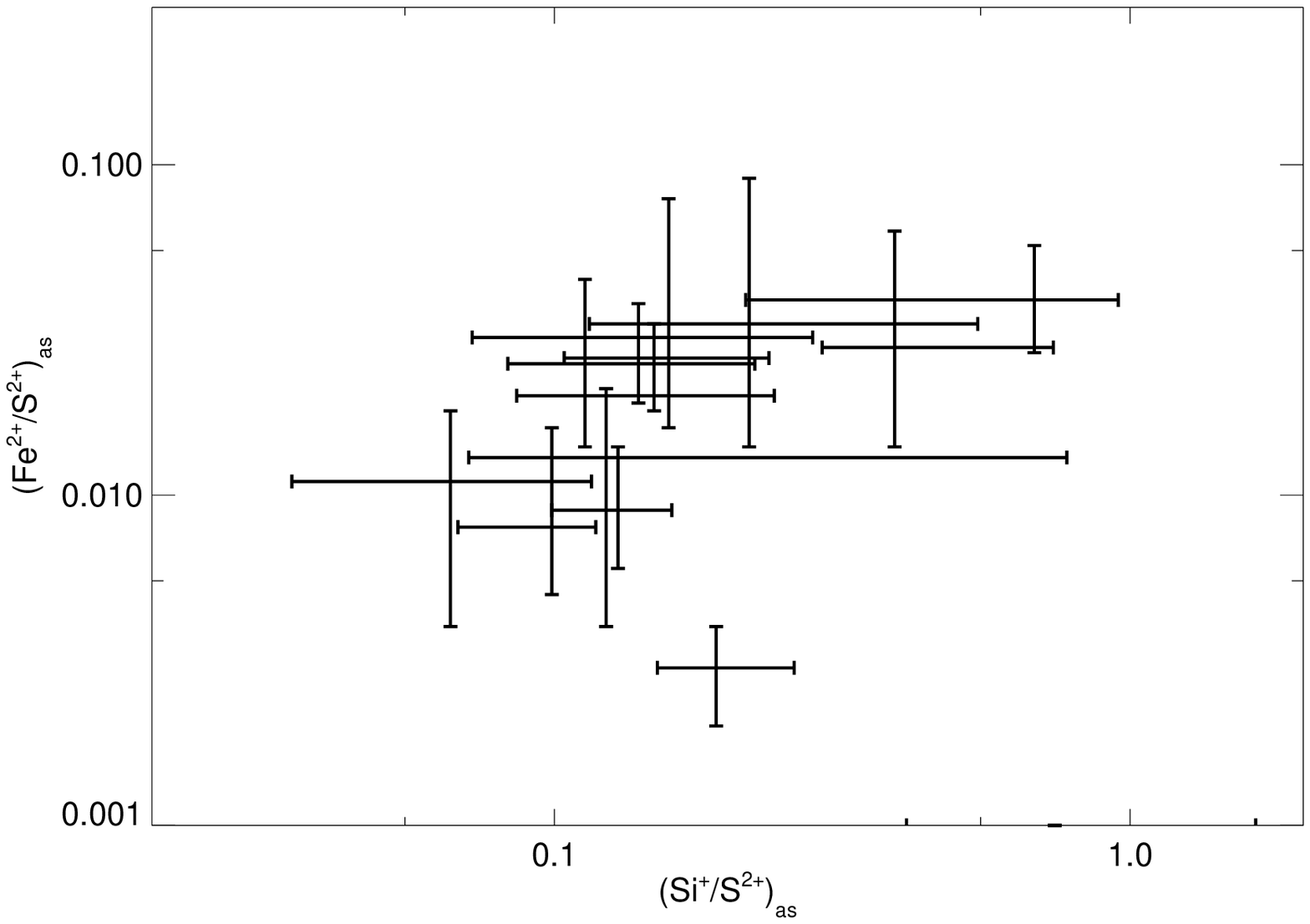}
\caption{(Fe$^{2+}$/S$^{2+}$)$_\mathrm{as}$ against (Si$^{+}$/S$^{2+}$)$_\mathrm{as}$.\label{FeIII23SiII35SuIII33_figure}}
\end{figure}

Table~\ref{FeIII23SiII35SuIII33_table} summarizes the derived (Fe$^{+}$/Si$^{+}$)$_\mathrm{as}$, (Si$^{+}$/S$^{2+}$)$_\mathrm{as}$, and (Fe$^{2+}$/S$^{2+}$)$_\mathrm{as}$ by assuming that \feii\ 26\um\ and \siii\ 35\um\ emissions come from the ionized gas as well as \feiii\ 23\um\ and \suiii\ 33\um, and that the abundance is constant within one star-forming region.  Those are plotted in Figs.~\ref{FeII26SiII35_figure} and \ref{FeIII23SiII35SuIII33_figure}.  In S171, (Fe$^{2+}$/S$^{2+}$)$_\mathrm{as}$ is estimated to be 1--2\% of the solar abundance.

Similarly to \siii\ 35\um, the \feii\ 26\um\ emission correlates with the H$_2$ 28\um\ and it has a negative correlation with the \suiii\ 33\um\ emission.  Again, non-uniform sampling over two molecular cloud regions of the present observations has to be taken into account in the interpretation.  With assuming that all the \feii\ 26\um\ emission comes from the PDR gas, we estimate (Fe$^{+}$)$_\mathrm{as}$ using PDR properties derived in \citet{Okada03} for each cloud.  \citet{Okada03} show that positions in the cloud near the exciting star have the radiation field strength ($G_0$) in units of the solar neighborhood value ($1.6\times 10^{-6}$\,W\,m$^{-2}$) of about 250--1200 and the gas density of the PDR ($n$) of 30--540\cc.  PDR models by \citet{Kaufman06} indicate \feii\ 26\um/H$_2$ 28\um $\sim 0.01$--0.04 for these parameters with the abundance in Table~\ref{abundance_table}.  Positions in the cloud far from the exciting star have $G_0=160$--420 and $n=850$--1300\cc, and the PDR models indicate \feii\ 26\um/H$_2$ 28\um $= 0.007$--0.017.  The average observed ratio is $0.23$ and $0.18$ for both clouds, respectively.  These suggest (Fe$^{+}$)$_\mathrm{as}=0.03$--$0.13$ and 0.06--0.15, respectively, by simple scaling.  On the other hand, if assuming that all the \feii\ 26\um\ comes from the ionized gas, the ratio to \nii\ 122\um\ observed by {\it ISO} indicates (Fe$^+$/N$^+$)$_\mathrm{as}=0.005$--$0.01$ and $0.01$--$0.05$ for both clouds, respectively.

We estimate the electron density from \suiii\ 18.7\um/33\um\ to be $170$\cc\ for an average, and less than $540$\cc\ for all the observed positions.  The presence of a diffuse \hii\ region in S171 is suggested by the electron density of $< 150$\cc\ and $<70$\cc\ derived from the \oiii\ 52\um/88\um\ and \nii\ 122\um/\cii\ 158\um, respectively \citep{Okada03}.  The present results show that the \suiii\ emitting gas has an electron density similar to that of \oiii\ and \nii\ and \cii\ emitting gas.

\subsubsection{G333.6-0.2}

The \suiii\ 33\um\ intensities reported by {\it KAO} observations at N1 (s4--s7), N2 (s8--s11), and N3 (s12--s15) are ($2850\pm 50$), ($990\pm 70$), and ($590\pm 50$)$\times 10^{-8}$\,W\,m$^{-2}$\,sr$^{-1}$, respectively, and \suiii\ 18.7\um\ intensities by {\it KAO} at N1 and N2 are ($2410\pm 240$) and ($550\pm 100$)$\times 10^{-8}$\,W\,m$^{-2}$\,sr$^{-1}$, respectively \citep{Simpson04}.  The observed \suiii\ 18.7\um\ intensities of the averaged spectra by {\it Spitzer} do not always agree with those by {\it KAO} within $1\sigma$, which may be attributed to the difference in the aperture size between the IRS ($4.7\arcsec\times 11.3\arcsec$) and {\it KAO} ($60\arcsec$) observations.  Table~\ref{lineSH} shows that the \suiii\ 18.7\um\ intensity within one mapping set (e.g. s4--s7) differs largely, and the discrepancy between {\it Spitzer} and {\it KAO} is not larger than these scatters within one mapping set.  However, the \suiii\ 33\um\ intensities by the IRS are at least by 30--40\% smaller than those by {\it KAO} and these can not simply be attributed to the difference of the aperture.  The cause is unknown at present.  The derived density from the extinction-corrected \suiii\ 18.7\um/33\um\ by the IRS ranges 700--2200\cc\ at s0--s3, 600--3000\cc\ at s4--s7, 600--1300\cc\ at s8--s11, and 300--1600\cc\ at s12--s15.  

The good correlation between \siii\ 35\um\ and \suiii\ 33\um\ (Fig.~\ref{SIIIH2SiII35} and Table~\ref{corr_coeff})  indicates that the major fraction of the \siii\ 35\um\ emission comes from the ionized gas.  The sampling of the present observations is rather uniform comparing to the case of S171 (Fig.~\ref{G333obspos}) and should have less effects from local conditions.  We use the \nii\ emission observed by {\it KAO} \citep{Simpson04} and derive (Si$^{+}$/N$^{+}$)$_\mathrm{as}=0.20$ ($\pm 0.05$), $0.31$ ($\pm 0.10$), and $0.13$ ($+0.83/-0.06$) for s(4--7), s(8--11), and s(12--15), respectively.  These are similar to (Si$^+$/S$^{2+}$)$_\mathrm{as}$ (Table~\ref{FeIII23SiII35SuIII33_table}).  For s(12--15), only the \nii\ 205\um\ emission has been detected by {\it KAO} and we use \nii\ 122\um\ emission for the other two positions.  The \siii\ 35\um/\nii\ 205\um\ ratio is sensitive to the density, thus the derived abundance ratio has a large uncertainty.  For s(4--7) and s(8--11), the derived (Si$^{+}$/N$^{+}$)$_\mathrm{as}$ is in the same order of magnitude as in S171 \citep{Okada03}.  Although the major fraction of the \siii\ 35\um\ emission is considered to come from the ionized gas as mentioned above, the contribution from the PDR should not be negligible.  In fact, the \siii\ 35\um/\nii\ 122\um\ ratio in the center of G333.6-0.2 \citep{Colgan93} indicates the Si abundance of over 100\% solar and thus part of \siii\ 35\um\ emission should come from the PDR.  We compare the observed \siii\ 35\um/H$_2$ S(0) 28\um\ with the PDR model by \citet{Kaufman06} for s8 and s12--s15, where H$_2$ 28\um\ emission has been detected.  PDR properties at the center of G333.6-0.2 have been derived to be $G_0=10^5$ and $n=4\times 10^4$\cc\ \citep{Colgan93}.  If the PDR at s8 and s12--s15 has the same $G_0$ and $n$ as those at the central region, the observed \siii\ 35\um/H$_2$ 28\um\ is consistent with the model with the abundance in Table~\ref{abundance_table}, i.e. (Si$^+$)$_\mathrm{as}=0.05$.  Those observed regions are far from the central region (Fig.~\ref{G333obspos}) and $G_0$ should be decreased with the square of the distance from exciting sources.  If $G_0$ decreases to $10^3$, the observed \siii\ 35\um/H$_2$ S(0) 28\um\ ratio becomes larger than the model prediction by a factor of 9--23, which corresponds to (Si$^+$)$_\mathrm{as}=0.5$--1.2.  

We estimate the (Fe$^{2+}$/S$^{2+}$)$_\mathrm{as}$ to be $\sim 0.03$ in G333.6-0.2 (Table~\ref{FeIII23SiII35SuIII33_table}).  Even if we assume that all the \feii\ 26\um\ emission comes from the ionized gas, the ratio to \nii\ emission observed by {\it KAO} indicates (Fe$^{+}$/N$^{+}$)$_\mathrm{as} < 0.05$.  The \feii\ 26\um/H$_2$ S(0) 28\um\ ratio in the PDR changes rapidly with $G_0$.  Thus it is difficult to obtain a useful constraint on (Fe$^{+}$)$_\mathrm{as}$ in the PDR.

\subsubsection{$\sigma$ Sco}

In the $\sigma$ Sco region, the \siii\ 35\um\ and \feii\ 26\um\ emission was detected for the first time.  The derived \siii\ 35\um\ intensities are compatible with the upper limit estimated by {\it ISO} observations \citep{Okada06}.  H$_2$ 28\um\ emission was detected whereas \citet{deGeus90} did not detect the CO line emission around $\sigma$ Sco.  \siii\ 35\um\ emission shows a negative correlation with the \suiii\ 33\um\ emission and the correlation coefficient indicates a good correlation of the \siii\ 35\um\ with the H$_2$ 28\um\ emission.  It should be noted that the uncertainty of the intensities are relatively larger than other targets (Fig.~\ref{SIIIH2SiII35}) and the sampling of the observed positions is not uniform along the distance from the exciting star as in S171.  \citet{Okada06} examined physical properties of both ionized and PDR gas in the $\sigma$ Sco region from far-infrared line and continuum emissions as $n_e<20$\cc\ for the ionized gas, $G_0=200$--1200 for the PDR gas at p7, p9, and p10, and $n=20$--50\cc\ for the PDR gas at p7.  If all the \siii\ 35\um\ emission comes from the ionized gas, the ratio to \nii\ 122\um\ observed by {\it ISO} indicates (Si$^{+}$/N$^+$)$_\mathrm{as} \sim 0.08$--0.25.  On the other hand, the \siii\ 35\um/H$_2$ S(0) 28\um\ indicates (Si$^+$)$_\mathrm{as}$ to be 0.2--1.6 if all the \siii\ 35\um\ emission comes from the PDR gas. 

\feiii\ 23\um\ has not been detected at any observed position and \feii\ 26\um\ has been detected only at s0--s4, the farthest position from the exciting star.  Assuming the PDR gas for the origin of \feii\ 26\um, (Fe$^+$)$_\mathrm{as}\sim 0.03$--0.22, and assuming the ionized gas origin, (Fe$^+$/N$^+$)$_\mathrm{as} \sim 0.03$--0.06.

\subsubsection{NGC~1977}

In NGC~1977, the \ariii\ 22\um, \feiii\ 23\um, and \suiii\ 33\um\ emissions, which come from the ionized gas, were not detected.  This is consistent with that all the observed positions are located outside the ionization front on the edge-on structure.  Thus \siii\ 35\um\ and \feii\ 26\um\ emissions are considered to come from the PDR.  At s0--s3, the \siii\ 35\um\ intensity detected by {\it KAO} is reported to be ($13\pm 4$)$\times 10^{-8}$\,W\,m$^{-2}$\,sr$^{-1}$ \citep{YoungOwl02}, which is weaker, but consistent with the intensity obtained by the IRS within $3\sigma$.

\begin{figure}
\epsscale{1.0}
\plotone{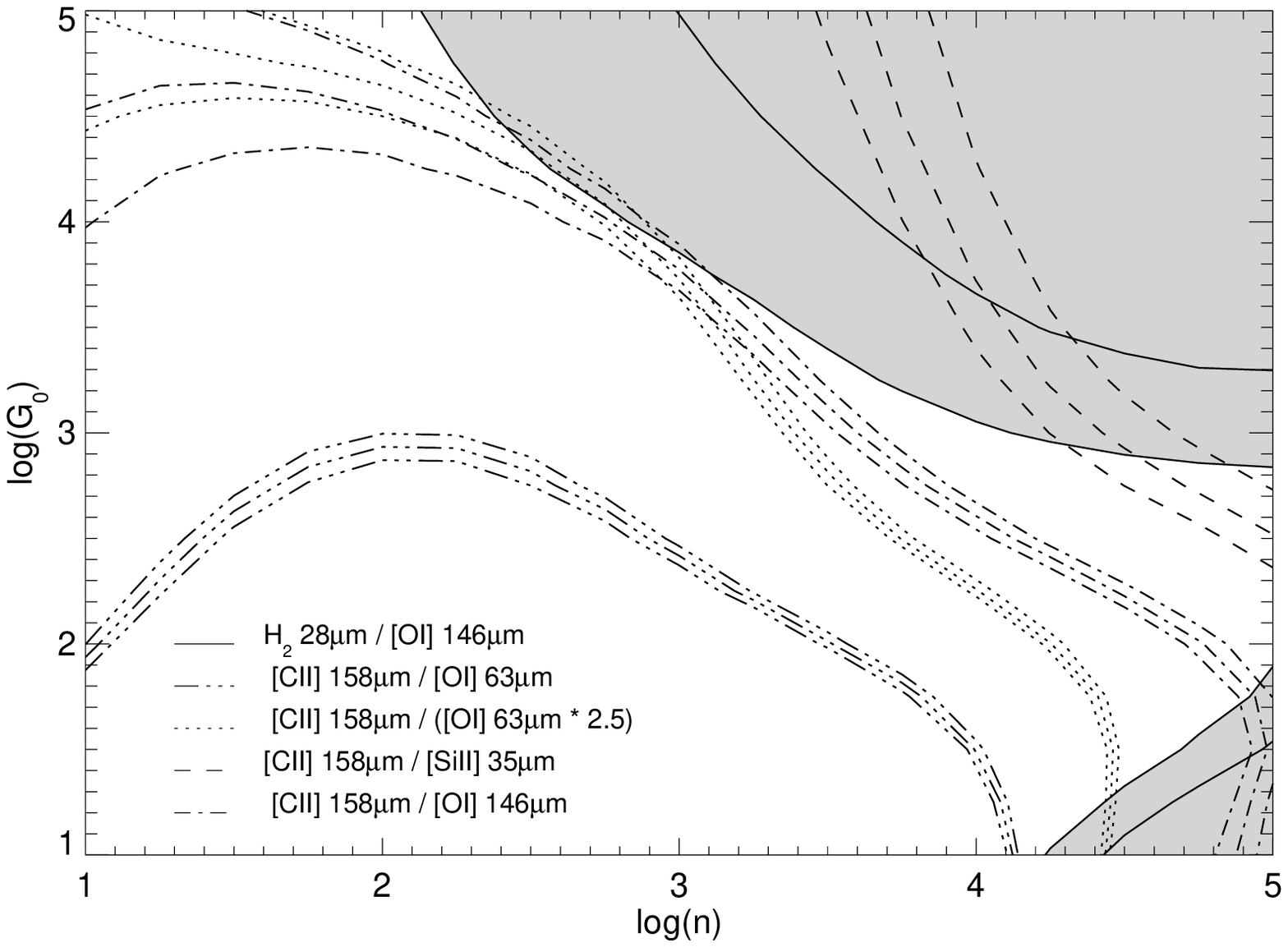}
\caption{PDR model parameters $G_0$ and $n$ by \citet{Kaufman06} that satisfy the observed line ratios among \cii\ 158\um, \oi\ 63\um, 146\um, \siii\ 35\um, and H$_2$ 28\um.  For \cii\ 158\um/\oi\ 63\um, the ratio if the intrinsic \oi\ 63\um\ line emission is attenuated by a factor 2.5 due to the optical depth effect is also shown by the dotted lines. $+1\sigma$ lines are always lower than the other two (larger ratios indicate lower $G_0$).  \cii\ 158\um, \oi\ 63\um, 146\um, and \siii\ 35\um\ intensities are from \citet{YoungOwl02}.  For H$_2$ 28\um\ observed by the IRS, the error represents the intensity range among s0--s4.  For the ratio of H$_2$ 28\um\ to \oi\ 146\um, the shaded regions satisfy the observed results.  \label{NGC1977_KAO}}
\end{figure}

\citet{YoungOwl02} have derived the PDR properties $G_0\sim 5000$ and $n=2\times 10^3$--$3\times 10^4$\cc\ from the observed intensities of \cii\ 158\um, \siii\ 35\um, and \oi\ 63\um\ at the \cii\ peak position corresponding to s0--s3.  We reanalyze their data of \oi\ 63\um, 146\um, \cii\ 158\um\ and \siii\ 35\um, together with H$_2$ S(0) 28\um\ detected by the IRS using the latest PDR model by \citet{Kaufman06}.  We plot the ratio of the observed those line intensities on the $n$--$G_0$ plane of the model in Fig.~\ref{NGC1977_KAO}.  It shows that any combinations of $G_0$ and $n$ can not reproduce the observed line ratios.  The discrepancy can be reduced if \oi\ 63\um\ is optically thick in overlapping clouds toward the line-of-sight and Si in gas phase is larger than assumed in the PDR model as in S171 and $\rho$ Oph \citep{Okada03, Okada06}.  If we assume that \oi\ 63\um\ emission is intrinsically stronger than the observed intensity by a factor of 2.5, the line ratios except for \cii\ 158\um/\siii\ 35\um\ in Fig.~\ref{NGC1977_KAO} meet at $G_0\sim 5000$ and $n\sim 10^3$\cc.  Then (Si$^+$)$_\mathrm{as}$ should be $\sim 0.16$, 3 times larger than Si abundance assumed in the PDR model (Table~\ref{abundance_table}), to account for all the line ratios.  As mentioned above, the \siii\ 35\um\ intensity observed by the IRS is larger than that by {\it KAO} by a factor of 1.9.  If we adopt the intensity by the IRS, (Si$^+$)$_\mathrm{as}$ should be 0.3.  In s4--s7 and s8--s11, \siii\ 35\um/H$_2$ 28\um\ is lower than that in s0--s3 (Table~\ref{lineLH}), which indicates lower values of $G_0$ and/or $n$, or (Si$^+$)$_\mathrm{as}\sim 0.1$.


\siii\ 35\um/\feii\ 26\um\ is 10--17 for all the positions.  Since the model prediction of the ratio is 20--25 for $G_0\sim 5000$ and $n\sim 10^3$\cc, (Fe$^+$)$_\mathrm{as}$ should be $\sim 0.02$--$0.04$, 4--8 times larger than Fe abundance in the model (Table~\ref{abundance_table}) if (Si$^+$)$_\mathrm{as}=0.16$.

\subsection{Group (b)}

For most group (b) targets, the \siii\ 35\um\ emission shows better correlation with \suiii\ 33\um\ than with H$_2$ 28\um\ (Fig.~\ref{SIIIH2SiII35} and Table~\ref{corr_coeff}).  By assuming that all the \siii\ 35\um\ emission comes from the ionized gas, we derive (Fe$^+$/Si$^+$)$_\mathrm{as}$ and (Si$^+$/S$^{2+}$)$_\mathrm{as}$ (Table~\ref{FeIII23SiII35SuIII33_table}).  (Fe$^{2+}$/S$^{2+}$)$_\mathrm{as}$ are also shown in Table~\ref{FeIII23SiII35SuIII33_table}.  In contrast to Si$^+$, (Fe$^{2+}$/S$^{2+}$)$_\mathrm{as}$ is below 0.1 even taking account of errors in all the 10 regions.

For G32.797+0.192 and G351.467-0.462, {\it ISO}/SWS and LWS single position observations have been reported by \citet{Peeters02} as IRAS18479-0005 and IRAS17221-3619, respectively (see Table~\ref{lineLH}).  We apply the extended source correction to line fluxes by \citet{Peeters02} and convert them to the surface brightness.  The obtained \suiv\ 10.5\um, \neii\ 12.8\um, \suiii\ 33\um, and \siii\ 35\um\ intensities of IRAS17221-3619 are consistent of those of G351.467-0.462 s0 within $2\sigma$, whereas \suiii\ 33\um\ and \siii\ 35\um\ intensities of IRAS18479-0005 are much smaller than those of G32.797+0.192 s1.  Table~\ref{lineLH} shows the large intensity gradient in G32.797+0.192, so this discrepancy may be due to the difference of the aperture between the SWS and IRS.  \siii\ 35\um/\nii\ 122\um\ with {\it ISO} observations indicates (Si$^+$/N$^{+}$)$_\mathrm{as}=0.19$--0.29 for IRAS17221-3619 (G351.467-0.462 s0), which is consistent with (Si$^+$/S$^{2+}$)$_\mathrm{as}$ (Table~\ref{FeIII23SiII35SuIII33_table}).  With assuming $A_V=60$~mag, (Si$^+$/N$^{+}$)$_\mathrm{as}$ should increase by a factor of 2, still consistent with (Si$^+$/S$^{2+}$)$_\mathrm{as}$ within the uncertainty.  \nii\ 122\um\ emission has not been detected in IRAS18479-0005 (G32.797+0.192 s1).

To derive the gas-phase abundance in the PDR, we need observations of cooling lines such as \oi\ 146\um\ and the far-infrared continuum in the PDR to constrain $n$ and $G_0$.  Since \siii\ 35\um/H$_2$ 28\um\ and \feii\ 26\um/H$_2$ 28\um\ changes rapidly with these PDR parameters, it is difficult to constrain (Si$^+$)$_\mathrm{as}$ and (Fe$^+$)$_\mathrm{as}$ using only H$_2$ 28\um.  For G32.797+0.192 and G351.467-0.462, we analyze \oi\ 63\um, 146\um, \cii\ 158\um, and \siii\ 35\um\ emission lines as well as far-infrared continuum emissions observed by {\it ISO} in the same manner as \citet{Okada03,Okada06}.  We fit the continuum emission of the LWS spectra with a graybody radiation of the $\lambda^{-1}$ emissivity to derive dust temperature, from which we estimate $G_0$ (Table~\ref{PDRISO}).  For both regions, no PDR models account for the derived $G_0$ and the observed \oi\ 63\um/FIR, \oi\ 146\um/FIR, and \cii\ 158\um/FIR, where FIR is the total far-infrared intensity.  We assume that PDR clouds overlap along the line-of-sight, with each cloud being $\tau=\infty$ for \oi\ 63\um, $\tau=1$ for \cii\ 158\um, and optically thin for \oi\ 146\um\ and continuum emissions, and estimate the degree of overlapping from $Z=\mathrm{FIR(obs)/FIR(model)}$ as in \citet{Okada03,Okada06}.  The observed \oi\ 63\um, 146\um, \cii\ 158\um, and the continuum emissions are consistent with the PDR model by \citet{Kaufman06} with the parameters of Table~\ref{PDRISO}.  \siii\ 35\um/FIR will be consistent with these PDR properties if (Si$^+$)$_\mathrm{as}=0.8$--1.3 and 1.0--2.0 for two regions (Table~\ref{PDRISO}).  The extinction correction increases (Si$^+$)$_\mathrm{as}$.  The estimated Si abundance is almost solar and is much higher than the abundance estimated if the line originates from the ionized gas (Table~\ref{FeIII23SiII35SuIII33_table}), which indicates that the major origin of \siii\ 35\um\ is the ionized gas in these two regions.

\section{Discussion}

\subsection{Gas-phase abundance}

The analysis in the previous section derives the ionic abundance ratio as (Fe$^+$/Si$^{+}$)$_\mathrm{as}=0.005$--0.38, (Si$^+$/S$^{2+}$)$_\mathrm{as}=0.03$--1.7, (Si$^+$/N$^{+}$)$_\mathrm{as}=0.08$--0.4, (Fe$^{2+}$/S$^{2+}$)$_\mathrm{as}=0.002$--0.08, and (Fe$^+$/N$^{+}$)$_\mathrm{as}=0.005$--0.06 in the ionized gas, and (Si$^+$)$_\mathrm{as}=0.16$--1.6 and (Fe$^+$)$_\mathrm{as}=0.02$--0.22 in the PDR gas.  In this subsection we examine the uncertainty in interpretation of these ratios as the gas-phase elemental abundance of Si and Fe against the solar abundance.

For the PDR gas, models take into account chemical reactions and the balance of the ionization, and the estimated (Si$^+$)$_\mathrm{as}$ and (Fe$^+$)$_\mathrm{as}$ should correspond to the gas-phase elemental abundance of Si and Fe against the solar abundance in the first order approximation.  The uncertainties in the estimation of $n$ and $G_0$ affect the results.  According to \citet{Kaufman06}, \siii\ 35\um/\oi\ 146\um\ is less sensitive to $n$ and $G_0$.  It varies only 50\% in a range of $n=10$--$10^4$\cc\ and $G_0=10^2$--$10^4$.  On the other hand, \siii\ 35\um/H$_2$ 28\um\ and \feii\ 26\um/H$_2$ 28\um\ are much more sensitive to $n$ and $G_0$, especially for $n>10^4$\cc\ and $G_0>10^4$.  Therefore, it is difficult to examine the abundance from \siii\ 35\um/H$_2$ 28\um\ and \feii\ 26\um/H$_2$ 28\um\ except for nearby star-forming regions, where the PDR parameters have been well estimated.

For the ionized gas, we need to take into account the ionization structure and the ionization fraction to translate the ionic abundance ratio into the gas-phase elemental abundance ratio.  Since the ionization potential of Si and Fe is similar, (Fe$^+$/Si$^{+}$)$_\mathrm{as}$ is a good indicator of (Fe/Si)$_\mathrm{as}$.  The first ionization potential of N is similar to that of hydrogen, and the second ionization potential is quite high compared to that of Si (Table~\ref{IPtable}).  Thus (Si$^+$/N$^+$)$_\mathrm{as}$ is a lower limit of the gas-phase elemental abundance of (Si/N)$_\mathrm{as}$.  This is also true for Fe$^+$.  Since N is little depleted \citep{SavageSembach96}, (Si$^+$/N$^+$)$_\mathrm{as}$ and (Fe$^+$/N$^+$)$_\mathrm{as}$ indicate lower limits of the gas-phase abundance of Si and Fe against the solar abundance, respectively.

S is also considered to not be heavily depleted in the ISM \citep{SavageSembach96}.  Though the difference between the gas-phase S abundance and the solar abundance has been discussed for \hii\ regions \citep{MartinHernandez02}, starburst galaxies \citep{Verma03}, and planetary nebulae \citep{Pottasch06}, \citet{MartinHernandez02} reported that regions with low S abundance have high density of $n_e>10^4$\cc.  Thus we assume the solar abundance as gas-phase S abundance.  Since the second ionization potential of S is much larger than the ionization potential of Si (Table~\ref{IPtable}), the derived (Si$^+$/S$^{2+}$)$_\mathrm{as}$ can not be assumed to be equal to (Si/S)$_\mathrm{as}$.  In fact, (Si$^+$/S$^{2+}$)$_\mathrm{as}$ in $\sigma$ Sco is too large probably because the excitation source in this region is a B1 star.  On the other hand, (Si$^+$/S$^{2+}$)$_\mathrm{as}$ gives a similar value as (Si$^+$/N$^+$)$_\mathrm{as}$, 0.22 and 0.13--0.31, respectively, in G333.6-0.2, which has $N$(Lyc)$>10^{50}$ photons s$^{-1}$ as all the group (b) targets.  Therefore, we assume that (Si$^+$/S$^{2+}$)$_\mathrm{as}$ for group (b) targets roughly indicates the gas-phase Si abundance against solar.  Fe$^{2+}$ and S$^{2+}$ have similar ionization potentials, thus (Fe$^{2+}$/S$^{2+}$)$_\mathrm{as}$ should be a good estimate for (Fe/S)$_\mathrm{as}$.

Since Ne$^+$ has also similar ionization potential with that of S$^{2+}$ and Ne is less depleted \citep{SavageSembach96}, we examine (Si$^+$/Ne$^{+}$)$_\mathrm{as}$ and (Fe$^{2+}$/Ne$^{+}$)$_\mathrm{as}$ as indicators for Si and Fe abundance for three targets with SH observations.  Those give smaller value by a factor of $\sim 3$ as (Si$^+$/S$^{2+}$)$_\mathrm{as}$ and (Fe$^{2+}$/S$^{2+}$)$_\mathrm{as}$, respectively, i.e., (Ne$^+$/S$^{2+}$)$_\mathrm{as}\sim 3$ for all the three targets.  Possible reasons are the contribution of the high density gas, the depletion of S, or unusual abundance of Ne.  The relative intensities of \suiii\ 18\um, 33\um, and \nii\ 122\um\ (if available) emissions are consistent with low density gas of $n_e\lesssim 10^3$\cc.  (Ne$^+$/Ar$^{2+}$)$_\mathrm{as}$ derived from \neii\ 12.8\um/\ariii\ 22\um\ also shows large value of $\sim 3$.  In addition, \citet{Simpson04} shows a high abundance of Ne/H compared to S/H, N/H, and O/H in G333.6-0.2.  Thus we do not adopt the \neii\ 12.8\um\ emission as a reference.

We use the solar abundance (Table~\ref{abundance_table}) as a reference elemental abundance.  The uncertainty of the relative solar abundance between two elements affects the results.  The solar abundance by \citet{AG89}, \citet{GrevesseSauval98}, \citet{Holweger01}, and \citet{Lodders03} differ by $+25$\% for Si, $+66$\%$/-1$\% for Fe, and $+55$\% for S from those by \citet{Asplund05} (Table~\ref{abundance_table}).  However, this is the systematic uncertainty and the result that gas-phase Si abundance is much larger than that of Fe does not change.  There is also a gradient of the abundance along the galactocentric distance, and the relative abundance ratio of two elements is affected by the chemical evolution there.  Basically, Si and S are $\alpha$-elements, and thus the Si/S ratio is less sensitive to the metallicity.  However Fe atoms are formed mostly in type Ia supernovae and N requires supplementary sources other than type II supernovae such as intermediate mass stars \citep{Hou00}.  The ratios between these elements vary with the metallicity.  The abundance gradient in our Galaxy has been derived observationally.  \citet{Daflon04} derive the slope of the abundance gradient as -0.04 dex kpc$^{-1}$ for Si and S, and -0.046 dex kpc$^{-1}$ for N from observations of young OB stars.  \citet{Chen03} show that the slope for Fe is -0.024  dex kpc$^{-1}$ from open clusters with ages less than 0.8 Gyr.  A simple extrapolation indicates that the solar abundance ratio of Si/N, Fe/N, and Fe/S will decrease by 0.048, 0.176, and 0.128 dex at the Galactic center, respectively.  The most affected ratio is Fe/N, which would result in an overestimate of the gas-phase Fe abundance by about 50\% at the Galactic center.  All the group (a) samples for which we derive the Fe/N abundance are, however, located within $\sim 3$~kpc.  Fe/S would result in an overestimate of the gas-phase Fe abundance only by about 30\% even at the Galactic center.  Thus the uncertainty of the reference abundance does not affect the gas-phase Si and Fe abundance discussed in the next subsection.  On the other hand, \citet{Knauth06} show that the abundance ratio of N/O derived from stars within $\sim 500$~pc of the Sun is larger than that from distant stars by $\sim 50$\%.  $\sigma$ Sco is within $\sim 500$~pc, thus Si$^+$/N$^+$ and Fe$^+$/N$^+$ have uncertainty of $+50$\%.

The fact that \siii\ 35\um\ and \feii\ 26\um\ emissions originate both from the ionized and PDR gas makes the interpretation more complicated.  \citet{Kaufman06} estimated the fraction of the intensity of \siii\ 35\um\ and \feii\ 26\um\ from PDRs and \hii\ regions for $n_e=1$--$10^3$\cc, $N$(Lyc)$=10^{49}$ and $10^{51}$ s$^{-1}$, for metallicity $Z=1$ and 3, which meet physical properties of the most targets in the present study.  As a result, \feii\ 26\um\ emission is dominated by the PDR in all but one model of $n_e=1$\cc\ and $N$(Lyc)$=10^{49}$, whereas whether \siii\ 35\um\ emission is dominated by the PDR or \hii\ regions depends on the parameters.  With typical parameters of $Z=1$, $n_e=100$\cc, and $N$(Lyc)$=10^{51}$, $I_\mathrm{SiII}$(PDR)/$I_\mathrm{SiII}$(\hii)$\sim 4$, but it depends rapidly on the metallicity and the electron density.  \citet{Abel05} self-consistently calculated the thermal and chemical structure of an \hii\ region and PDR with CLOUDY, and investigated the percentage of the \siii\ 35\um\ line that is produced in the \hii\ region with various physical properties.  For the effective temperature of the exciting source $T_\mathrm{eff}=38000$~K, $n_e<70$\cc, and $G_0\sim 10^2$--$10^3$, which are properties in S171, the contribution of the \siii\ 35\um\ emission from the ionized gas is estimated to be $\gtrsim 30$\%.  When $n_e$ becomes lower, or $T_\mathrm{eff}$ becomes higher, or $G_0$ becomes lower, the contribution from the ionized gas increases.  The origin of the \feii\ 26\um\ may be examined by the \feii\ 18\um\ emission.  Since this upper level transition has the excitation energy of $2700$~K, the contribution from the PDR should be negligible for \feii\ 18\um.  We estimate the upper limit of the fraction of the \feii\ 26\um\ from the ionized gas from the observed upper limit of the \feii\ 18\um\ (Table~\ref{ionizedFeII26}).  Though strong constraints can not be obtained, the contribution from the PDR gas is not negligible in part of the observed positions.  Observations of \feii\ 18\um\ with a higher sensitivity in future missions will be useful to investigate ionized gas contributions.

\subsection{Si- and Fe-bearing dust grains}

Table~\ref{FeIII23SiII35SuIII33_table} and Fig.~\ref{FeII26SiII35_figure} show that (Fe$^+$/Si$^{+}$)$_\mathrm{as}$ is less than 0.1 in most regions with assuming that \siii\ 35\um\ and \feii\ 26\um\ emissions come from the ionized gas.  It is clearly shown that Fe is highly depleted than Si.  The analysis in the previous section shows that the gas-phase Si abundance indicated by (Si$^+$/S$^{2+}$)$_\mathrm{as}$ and (Si$^+$/N$^+$)$_\mathrm{as}$ is 3\%--100\% of the solar abundance in the ionized gas in a wide range of the exciting stars, i.e. from $\sigma$ Sco to giant \hii\ regions with $N$(Lyc)$>10^{50}$ photons s$^{-1}$.  Together with the result of \citet{Okada03,Okada06}, several PDRs also have similar gas-phase Si abundance.  On the other hand, the gas-phase Fe abundance indicated by (Fe$^{2+}$/S$^{2+}$)$_\mathrm{as}$ and (Fe$^+$/N$^+$)$_\mathrm{as}$ is $<8$\% of the solar abundance in all the observed regions (Fig.~\ref{FeIII23SiII35SuIII33_figure}).  From UV observations, the gas-phase abundance of Si and Fe in cool interstellar clouds are 5\% and 0.5\% solar, and those in warm interstellar clouds are 30\% and 6\% solar \citep{SavageSembach96}.  In diffuse interstellar clouds probed by UV observations, shock waves by supernovae destroy dust grains.  The gas-phase abundance of several elements including Si and Fe shows a trend to increase with the velocity of the clouds, indicating that shocks that accelerate clouds may be a dominant process for dust destruction \citep{Fitzpatrick96,Cowie78}.  The gas-phase Si and Fe abundance in star-forming regions discussed in the present paper is in a range similar to that of warm interstellar clouds.  However, dust grains in star-forming regions should originate from dense cool clouds, where Si and Fe are highly depleted \citep{Joseph86}, and they are processed by UV radiation or stellar winds from newly born massive stars in a different way from grains in warm interstellar clouds.  

It is generally accepted that interstellar dust grains consist of amorphous silicates and some form of carbonaceous materials \citep{Draine03a}.  Si, Mg and Fe can be major constituents of silicate, but UV observations indicate that they show different depletion patterns; Si atoms return to the gas phase most easily, Mg is depleted more than or similarly to Si, and Fe atoms tend to insistently remain in dust grains \citep{Sofia94, Fitzpatrick96, Jones00, Cartledge04, Cartledge06}.  These indicate that silicates contain primarily Mg and most or all Fe atoms are in other components of dust such as metal or oxides.  Although the trend that Si atoms return to the gas phase more easily than Fe is the same in star-forming regions discussed in the present paper as in diffuse ISM, silicate should survive generally in the ionized regions since it has a large binding energy of $\sim 5$~eV \citep{Tielens98}.  The present study indicates that 3--100\% of Si atoms and $<22$\% of Fe atoms are included in dust grains which are destroyed easily in \hii\ regions, probably by the UV radiation.  The presence of mantles which release Si atoms into the gas phase by photodesorption by UV photons \citep{Walmsley99} or organometallic complexes such as PAH cluster \citep{Marty94,Klotz95} is suggested.  Another possibility is that a fraction of Si and Fe atoms are in very small grains.  
\citet{Rodriguez02} observed optical \feiii\ and \feii\ lines in bright Galactic \hii\ regions and found the Fe abundance against O in the ionized gas to be between 2\% and 30\% of solar.  The depletion of Fe is shown to correlate with the ionization degree, which indicates that dust grains are in small sizes with $a\sim 10$\AA, which can be significantly high temperature that leads to evaporation after the absorption of one or a few energetic photons.  The depletion is also shown to correlate with the color excess, coming from foreground dust grains.  They suggested that both correlations seem to require the presence of dust grains differing from classical grains in the respects of their small sizes or loose structure.  A fraction of those grains are destroyed in the ionized gas and surviving ones would be pushed outside the ionized region, which produce most of the color excess.

\section{Summary}

We observed 14 Galactic star-forming regions with the IRS on board {\it Spitzer}.  \siii\ 35\um\ is detected at all the observed positions and \feii\ 26\um\ and \feiii\ 23\um\ emissions were detected at more than half of the mapping positions of most targets.  The high sensitivity of the IRS enables to detect these emission lines from iron ions widely in low density star-forming regions for the first time.  For three group (a) targets, S171, G333.6-0.2, and $\sigma$ Sco, we provide (Si$^+$/N$^+$)$_\mathrm{as}\sim 0.08$--0.4 and (Fe$^+$/N$^+$)$_\mathrm{as}\sim 0.005$--0.01, $<0.05$, and 0.03--0.06, respectively, assuming that all the \siii\ 35\um\ and \feii\ 26\um\ emission originate from the ionized gas.  For both group (a) and (b) targets, we derived (Si$^+$/S$^{2+}$)$_\mathrm{as}$ and (Fe$^{2+}$/S$^{2+}$)$_\mathrm{as}$ to be 0.03--1 and $<0.08$, respectively.  We have suggested that those indicate roughly the elemental gas-phase abundance ratio of (Si/S)$_\mathrm{as}$ and (Fe/S)$_\mathrm{as}$ in most regions.  Together with the estimates in the PDR gas, the results indicate that 3--100\% of Si atoms and $<22$\% of Fe atoms are included in dust grains which are destroyed easily in the ionized or PDR gas in \hii\ regions by the UV radiation, or reside in mantles which are photodesorbed by UV photons, organometallic complexes, or in small grains.

\acknowledgments
This work is based on observations made with the {\it Spitzer Space Telescope}, which is operated by the Jet Propulsion Laboratory, California Institute of Technology under a contract with NASA.  We thank T. L. Roellig for his help in the reduction of the IRS data and providing data needed for the reduction.  We also thank T. Kamizuka for providing some programs to reduce the IRS data.  This work is supported in part by a Grant-in-Aid for Science Research from the Japan Society for the Promotion of Science (JSPS).

\clearpage

\begin{deluxetable}{cccccc}
\tabletypesize{\scriptsize}
\tablecaption{Summary of the observed targets.\label{targetsummary}}
\tablewidth{0pt}
\tablehead{
\colhead{Target} & \colhead{module} & \colhead{t\tablenotemark{a} [sec]} & \colhead{$N$(Lyc)\tablenotemark{b}[sec $^{-1}$]} & \colhead{$n_e$\tablenotemark{c} [\cc]} & \colhead{Reference}
}
\startdata
\multicolumn{6}{l}{group a}\\
\tableline
S171 & SH/LH & 240/(1200 or 480) & $8.94\times 10^{48}$ & $<70$ & 1,2\\
G333.6-0.2 & SH/LH & 30/132 or 162/42 & $2.69\times 10^{50}$ & $<210$ & 3,4\\
$\sigma$ Sco & LH & 600 & $3.5\times 10^{45}$ & $<20$ & 5,6\\
NGC~1977 & LH & 12 & $4\times 10^{46}$ & $102$ & 7,8\\
\tableline
\multicolumn{6}{l}{group b}\\
\tableline
G0.572-0.628 & LH & 18 & $1.1\times 10^{50}$ & 117 & 6\\
G3.270-0.101 & LH & 216 & $3.0\times 10^{50}$ & 31 & 6\\
G4.412+0.118 & LH & 288 & $3.6\times 10^{50}$ & 38 & 6\\
G8.137+0.228 & LH & 36 & $3.0\times 10^{50}$ & 160 & 6\\
G32.797+0.192 & LH & 180 & $1.1\times 10^{50}$ & 102 & 6\\
G48.930-0.286 & LH & 84 & $1.1\times 10^{50}$ & 117 & 6\\
G79.293+1.296 & LH & 48 & $1.4\times 10^{50}$ & 110 & 6\\
G347.611+0.2 & LH & 348 & $2.8\times 10^{50}$ & 57 & 6\\
G351.467-0.462 & SH/LH & 12/84 & $1.5\times 10^{50}$ & 64 & 6\\
G359.429-0.090 & LH & 126 & $2.9\times 10^{50}$ & 93 & 6\\
\enddata
\tablecomments{Reference: (1) \citet{Okada03}; (2) \citet{Angerhofer77}; (3) \citet{Colgan93}; (4) \citet{Conti04}; (5) \citet{Okada06}; (6) \citet{Baart80}; (7) \citet{Goss70}; (8) \citet{Shaver70}}
\tablenotetext{a}{On source integration time for each mapping position in second.}
\tablenotetext{b}{The number of Lyman continuum photon.}
\tablenotetext{c}{Electron density.}
\end{deluxetable}

\clearpage
\pagestyle{empty}
\begin{deluxetable}{cccllccccccc}
\tabletypesize{\scriptsize}
\rotate
\tablecaption{LH line intensities without extinction correction.\label{lineLH}}
\tablewidth{0pt}
\tablehead{
\colhead{Target} & \colhead{s\#} & \colhead{cf.\tablenotemark{a}} & \colhead{RA} & \colhead{DEC} & \multicolumn{7}{c}{Line intensities [$10^{-8}$\,W\,m$^{-2}$\,sr$^{-1}$]}\\
\colhead{} & \colhead{} & \colhead{} & \multicolumn{2}{c}{(J2000)} & \colhead{\ariii} & \colhead{\feiii} & \colhead{\feii} & \colhead{H$_2$} & \colhead{\suiii} & \colhead{\siii} & \colhead{\neiii}\\
\colhead{} & \colhead{} & \colhead{} & \colhead{} & \colhead{} & \colhead{22\um} & \colhead{23\um} & \colhead{26\um} & \colhead{28\um} & \colhead{33\um} & \colhead{35\um} & \colhead{36\um}
}
\startdata
\multicolumn{12}{l}{group a}\\
\tableline
S171 & 0 && 0h1m18.56s & 67d23m0.51s &  0.43 $\pm$ 0.03 &  0.27 $\pm$ 0.02 &  0.13 $\pm$ 0.02 &  0.70 $\pm$ 0.05 & 43.88 $\pm$ 0.22 &  5.86 $\pm$ 0.38 &  $<$ 0.72  \\
& 1 && 0h1m16.68s & 67d22m43.72s &  0.42 $\pm$ 0.03 &  0.26 $\pm$ 0.02 &  0.13 $\pm$ 0.03 &  0.75 $\pm$ 0.05 &  44.04 $\pm$ 0.22 &  6.45 $\pm$ 0.37 &  $<$ 0.64 \\
& 2 && 0h1m13.04s & 67d22m57.25s &  0.42 $\pm$ 0.03 &  0.26 $\pm$ 0.02 &  0.12 $\pm$ 0.02 &  0.78 $\pm$ 0.05 &  43.03 $\pm$ 0.24 &  6.27 $\pm$ 0.36 &  $<$ 0.58 \\\
& 3 && 0h1m14.91s & 67d23m14.07s &  0.42 $\pm$ 0.03 &  0.25 $\pm$ 0.01 &  0.11 $\pm$ 0.02 &  0.86 $\pm$ 0.05 &  42.20 $\pm$ 0.21 &  5.72 $\pm$ 0.35 &  $<$ 0.65 \\
& 0--3 & p8\tablenotemark{b} & \nodata & \nodata  &  0.42 $\pm$ 0.03 &  0.25 $\pm$ 0.02 &  0.12 $\pm$ 0.02 &  0.81 $\pm$ 0.05 &  43.33 $\pm$ 0.21 &  6.12 $\pm$ 0.36 &  $<$ 0.64 \\
& 4 && 0h1m9.08s & 67d22m36.03s &  0.46 $\pm$ 0.05 &  0.26 $\pm$ 0.03 &  0.11 $\pm$ 0.04 &  0.46 $\pm$ 0.05 &  37.15 $\pm$ 0.50 &  5.90 $\pm$ 0.14 &  1.06 $\pm$ 0.10 \\
& 5 && 0h1m5.62s & 67d22m35.84s &  0.46 $\pm$ 0.05 &  0.27 $\pm$ 0.03 &  0.14 $\pm$ 0.04 &  0.44 $\pm$ 0.05 &  37.46 $\pm$ 0.56 &  6.01 $\pm$ 0.14 &  1.01 $\pm$ 0.10 \\
& 6 && 0h1m5.57s & 67d23m0.83s &  0.47 $\pm$ 0.05 &  0.25 $\pm$ 0.03 &  0.15 $\pm$ 0.04 &  0.52 $\pm$ 0.06 &  39.09 $\pm$ 0.51 &  6.35 $\pm$ 0.17 &  0.98 $\pm$ 0.11 \\
& 7 && 0h1m9.04s & 67d23m1.03s &  0.45 $\pm$ 0.05 &  0.27 $\pm$ 0.03 &  0.12 $\pm$ 0.04 &  0.65 $\pm$ 0.06 &  39.80 $\pm$ 0.51 &  6.04 $\pm$ 0.17 &  1.07 $\pm$ 0.11 \\
& 4--7 & p9\tablenotemark{b} & \nodata & \nodata &   0.46 $\pm$ 0.05 &  0.25 $\pm$ 0.03 &  0.13 $\pm$ 0.04 &  0.50 $\pm$ 0.06 &  38.39 $\pm$ 0.52 &  6.06 $\pm$ 0.16 &  1.03 $\pm$ 0.10 \\
& 8 && 0h1m1.61s & 67d22m39.51s &  0.43 $\pm$ 0.03 &  0.21 $\pm$ 0.01 &  0.15 $\pm$ 0.02 &  0.37 $\pm$ 0.04 &  41.13 $\pm$ 0.36 &  7.41 $\pm$ 0.33 &  $<$ 0.50 \\
& 9 && 0h0m59.73s & 67d22m22.71s &  0.40 $\pm$ 0.03 &  0.22 $\pm$ 0.01 &  0.09 $\pm$ 0.02 &  0.29 $\pm$ 0.04 &  40.36 $\pm$ 0.28 &  6.80 $\pm$ 0.30 &  $<$ 0.42 \\
& 10 && 0h0m56.09s & 67d22m36.26s &  0.37 $\pm$ 0.03 &  0.27 $\pm$ 0.01 &  0.09 $\pm$ 0.01 &  0.34 $\pm$ 0.04 &  38.07 $\pm$ 0.26 &  6.52 $\pm$ 0.29 &  $<$ 0.41 \\
& 11 && 0h0m57.96s & 67d22m53.07s &  0.38 $\pm$ 0.03 &  0.20 $\pm$ 0.01 &  0.16 $\pm$ 0.02 &  0.39 $\pm$ 0.04 &  38.54 $\pm$ 0.29 &  7.78 $\pm$ 0.34 &  $<$ 0.52 \\
& 8--11 & p10\tablenotemark{b} & \nodata & \nodata  &  0.39 $\pm$ 0.03 &  0.22 $\pm$ 0.01 &  0.12 $\pm$ 0.02 &  0.33 $\pm$ 0.04 &  39.53 $\pm$ 0.28 &  7.08 $\pm$ 0.31 &  $<$ 0.47 \\
& 12 && 23h59m45.44s & 67d21m4.10s &  0.24 $\pm$ 0.03 &  0.25 $\pm$ 0.02 &  0.31 $\pm$ 0.02 &  1.13 $\pm$ 0.04 &  42.60 $\pm$ 0.24 &  15.33 $\pm$ 0.51 &  $<$ 0.42 \\
& 13 && 23h59m43.58s & 67d20m47.24s &  0.19 $\pm$ 0.03 & 0.22 $\pm$ 0.02 &  0.41 $\pm$ 0.03 &  1.23 $\pm$ 0.04 &  34.58 $\pm$ 0.23 &  15.69 $\pm$ 0.52 &  $<$ 0.56 \\
& 14 && 23h59m39.93s & 67d21m0.67s &  0.22 $\pm$ 0.04 &  0.19 $\pm$ 0.02 &  0.36 $\pm$ 0.02 &  0.75 $\pm$ 0.04 &  33.46 $\pm$ 0.18 &  13.64 $\pm$ 0.47 &  $<$ 0.44 \\
& 15 && 23h59m41.79s & 67d21m17.55s &  0.25 $\pm$ 0.03 &  0.22 $\pm$ 0.02 &  0.29 $\pm$ 0.02 &  0.98 $\pm$ 0.04 &  38.17 $\pm$ 0.23 &  14.54 $\pm$ 0.49 &  $<$ 0.43 \\
& 12--15 & p19\tablenotemark{b} & \nodata & \nodata  &  0.23 $\pm$ 0.03 &  0.22 $\pm$ 0.01 &  0.34 $\pm$ 0.02 &  1.00 $\pm$ 0.04 &  36.97 $\pm$ 0.21 &  14.84 $\pm$ 0.49 &  $<$ 0.44 \\
& 16 && 23h59m36.99s & 67d20m53.39s &  0.21 $\pm$ 0.03 &  0.12 $\pm$ 0.02 &  0.35 $\pm$ 0.02 &  0.71 $\pm$ 0.04 &  28.34 $\pm$ 0.19 &  10.93 $\pm$ 0.37 &  $<$ 0.42 \\
& 17 && 23h59m35.13s & 67d20m36.54s &  0.22 $\pm$ 0.03 &  0.09 $\pm$ 0.02 &  0.30 $\pm$ 0.01 &  2.48 $\pm$ 0.05 &  24.23 $\pm$ 0.14 &  10.53 $\pm$ 0.40 &  $<$ 0.48 \\
& 18 && 23h59m31.48s & 67d20m49.97s &  0.19 $\pm$ 0.03 &  0.13 $\pm$ 0.02 &  0.24 $\pm$ 0.02 &  2.92 $\pm$ 0.06 &  21.02 $\pm$ 0.15 &  8.99 $\pm$ 0.37 &  $<$ 0.44 \\
& 19 && 23h59m33.34s & 67d21m6.85s &  0.15 $\pm$ 0.03 &  0.11 $\pm$ 0.02 &  0.37 $\pm$ 0.03 &  1.44 $\pm$ 0.04 &  22.05 $\pm$ 0.14 &  11.54 $\pm$ 0.42 &  $<$ 0.39 \\
& 16--19 & p20\tablenotemark{b} & \nodata & \nodata  &  0.18 $\pm$ 0.03 &  0.11 $\pm$ 0.02 &  0.31 $\pm$ 0.02 &  1.86 $\pm$ 0.05 &  23.65 $\pm$ 0.14 &  10.42 $\pm$ 0.38 &  $<$ 0.41 \\
& 20 && 23h59m28.54s & 67d20m42.59s &  0.21 $\pm$ 0.03 &  0.11 $\pm$ 0.02 &  0.38 $\pm$ 0.03 &  1.61 $\pm$ 0.05 &  23.24 $\pm$ 0.19 &  10.66 $\pm$ 0.40 &  $<$ 0.42 \\
& 21 && 23h59m26.68s & 67d20m25.74s &  0.14 $\pm$ 0.02 &  0.09 $\pm$ 0.02 &  0.51 $\pm$ 0.02 &  2.73 $\pm$ 0.06 &  22.08 $\pm$ 0.21 &  15.47 $\pm$ 0.56 &  $<$ 0.67 \\
& 22 && 23h59m23.03s & 67d20m39.18s &  0.12 $\pm$ 0.03 &  0.08 $\pm$ 0.02 &  0.73 $\pm$ 0.04 &  2.91 $\pm$ 0.08 &  17.08 $\pm$ 0.20 &  20.96 $\pm$ 0.62 &  $<$ 0.94 \\
& 23 && 23h59m24.89s & 67d20m56.05s &  0.21 $\pm$ 0.03 &  0.07 $\pm$ 0.02 &  0.42 $\pm$ 0.02 &  1.35 $\pm$ 0.05 &  23.56 $\pm$ 0.16 &  11.68 $\pm$ 0.43 &  $<$ 0.52 \\
& 20--23 & p21\tablenotemark{b} & \nodata & \nodata  &  0.17 $\pm$ 0.02 &  0.08 $\pm$ 0.02 &  0.49 $\pm$ 0.02 &  2.18 $\pm$ 0.05 &  21.93 $\pm$ 0.15 &  14.56 $\pm$ 0.51 &  $<$ 0.65 \\

G333.6-0.2 & 0 && 16h22m5.48s & -50d5m40.17s  &  $<$ 4.22 &  12.61 $\pm$ 1.44 &  20.41 $\pm$ 1.57 &  $<$ 9.95 &  592.89 $\pm$ 15.50 &  452.53 $\pm$ 31.03 &  $<$ 63.66 \\  
& 1 && 16h22m4.35s & -50d5m23.40s   &  9.51 $\pm$ 2.27 &  19.87 $\pm$ 1.05 &  12.34 $\pm$ 1.80 &  $<$ 7.26 &  1272.78 $\pm$ 17.68 &  433.28 $\pm$ 24.93 &  $<$ 39.07 \\
& 2 && 16h22m6.53s & -50d5m9.75s	  &  29.06 $\pm$ 6.36 &  46.31 $\pm$ 6.14 &  $<$ 19.51 &  $<$ 18.78 &  2165.74 $\pm$ 24.89 &  484.31 $\pm$ 34.90 &  $<$ 68.37 \\	  
& 3 && 16h22m7.66s & -50d5m26.58s     &  $<$ 6.32 &  33.91 $\pm$ 2.50 &  23.94 $\pm$ 4.16 &  $<$ 18.79 &  1293.76 $\pm$ 27.70 &  564.37 $\pm$ 45.62 &  $<$ 88.70 \\	  
& 0--3 && \nodata & \nodata &  12.74 $\pm$ 3.40 &  28.50 $\pm$ 2.53 &  19.60 $\pm$ 3.61 &  $<$ 13.17 &  1304.98 $\pm$ 23.06 &  483.59 $\pm$ 32.50 &  $<$ 64.60 \\
& 4 && 16h22m9.73s & -50d4m49.39s  &  17.40 $\pm$ 3.33 &  $<$ 21.40 &  $<$ 10.76 &  $<$ 12.90 &  1141.71 $\pm$ 9.86 &  275.55 $\pm$ 6.50 &  $<$ 16.14 \\			     
& 5 && 16h22m10.97s & -50d5m5.48s  &  21.24 $\pm$ 3.78 &  34.70 $\pm$ 4.08 &  13.44 $\pm$ 4.38 &  $<$ 11.58 &  1549.81 $\pm$ 18.49 &  458.53 $\pm$ 10.23 &  $<$ 19.85 \\	     
& 6 && 16h22m8.89s & -50d5m20.45s  &  $<$ 16.94 &  $<$ 25.21 &  $<$ 20.56 &  $<$ 24.80 &  1132.36 $\pm$ 19.18 &  483.44 $\pm$ 11.56 &  $<$ 31.26 \\			     
& 7 && 16h22m7.65s & -50d5m4.45s	  &  59.63 $\pm$ 11.73 &  54.98 $\pm$ 8.67 &  $<$ 40.52 &  $<$ 36.21 &  2122.42 $\pm$ 55.59 &  516.86 $\pm$ 12.01 &  33.04 $\pm$ 10.41 \\     
& 4--7 & N1\tablenotemark{c} & \nodata & \nodata &  19.84 $\pm$ 5.96 &  22.22 $\pm$ 7.02 &  $<$ 28.80 &  $<$ 17.06 &  1329.90 $\pm$ 27.62 &  483.29 $\pm$ 9.54 &  $<$ 23.66 \\
& 8 && 16h22m9.72s & -50d3m49.55s  &  4.89 $\pm$ 0.74 &  4.02 $\pm$ 1.02 &  2.23 $\pm$ 0.62 &  $<$ 2.22 &  488.39 $\pm$ 4.65 &  178.10 $\pm$ 4.54 &  3.75 $\pm$ 1.07 \\	     
& 9 && 16h22m10.97s & -50d4m5.44s  &  4.63 $\pm$ 0.62 &  4.34 $\pm$ 0.89 &  $<$ 2.59 &  4.55 $\pm$ 1.40 &  528.11 $\pm$ 5.54 &  182.99 $\pm$ 4.53 &  $<$ 4.43 \\		     
& 10 && 16h22m8.89s & -50d4m20.42s &  $<$ 9.65 &  $<$ 8.36 &  $<$ 3.47 &  $<$ 5.22 &  699.30 $\pm$ 8.52 &  189.92 $\pm$ 4.60 &  $<$ 6.37 \\				     
& 11 && 16h22m7.65s & -50d4m4.43s  &  6.45 $\pm$ 1.50 &  4.64 $\pm$ 1.40 &  $<$ 3.48 &  $<$ 3.77 &  500.81 $\pm$ 5.39 &  166.20 $\pm$ 4.41 &  4.80 $\pm$ 1.11 \\	
& 8--11 & N2\tablenotemark{c} & \nodata & \nodata  &  6.31 $\pm$ 1.41 &  4.53 $\pm$ 1.32 &  $<$ 3.43 &  $<$ 3.56 &  536.16 $\pm$ 6.56 &  181.35 $\pm$ 4.51 &  $<$ 4.29 \\
& 12 && 16h22m9.72s & -50d2m49.54s &  2.35 $\pm$ 0.24 &  6.03 $\pm$ 0.22 &  2.75 $\pm$ 0.26 &  3.16 $\pm$ 0.26 &  336.14 $\pm$ 3.77 &  185.98 $\pm$ 4.51 &  $<$ 1.71 \\	     
& 13 && 16h22m10.97s & -50d3m5.45s &  3.08 $\pm$ 0.39 &  5.21 $\pm$ 0.26 &  2.38 $\pm$ 0.25 &  3.51 $\pm$ 0.37 &  402.81 $\pm$ 3.94 &  187.88 $\pm$ 4.81 &  $<$ 2.04 \\	     
& 14 && 16h22m8.89s & -50d3m20.42s &  2.29 $\pm$ 0.26 &  3.10 $\pm$ 0.32 &  1.57 $\pm$ 0.26 &  3.50 $\pm$ 0.39 &  306.58 $\pm$ 2.95 &  158.15 $\pm$ 4.07 &  $<$ 1.80 \\	     
& 15 && 16h22m7.65s & -50d3m4.41s  &  2.11 $\pm$ 0.23 &  4.35 $\pm$ 0.18 &  2.15 $\pm$ 0.21 &  3.17 $\pm$ 0.28 &  336.33 $\pm$ 2.92 &  208.05 $\pm$ 5.13 &  $<$ 1.36 \\	     
& 12--15 & N3\tablenotemark{c} & \nodata & \nodata  &  2.47 $\pm$ 0.24 &  4.54 $\pm$ 0.21 &  2.25 $\pm$ 0.25 &  3.32 $\pm$ 0.29 &  340.61 $\pm$ 3.17 &  188.98 $\pm$ 4.68 &  $<$ 1.\\

$\sigma$ Sco &  0 && 16h20m10.37s & -25d21m33.85s  &  $<$ 0.07 &  $<$ 0.10 &  0.11 $\pm$ 0.03 &  0.24 $\pm$ 0.03 &  0.55 $\pm$ 0.08 &  1.13 $\pm$ 0.09 &  $<$ 0.18 \\		
& 1 && 16h20m11.23s & -25d21m50.06s		  &  $<$ 0.08 &  $<$ 0.09 &  0.09 $\pm$ 0.03 &  0.25 $\pm$ 0.03 &  0.60 $\pm$ 0.09 &  1.13 $\pm$ 0.09 &  $<$ 0.17 \\		
& 2 && 16h20m9.74s & -25d22m4.66s		  &  $<$ 0.07 &  $<$ 0.09 &  0.11 $\pm$ 0.03 &  0.26 $\pm$ 0.03 &  0.58 $\pm$ 0.10 &  1.25 $\pm$ 0.10 &  $<$ 0.15 \\		
& 3 && 16h20m8.88s & -25d21m48.42s		  &  $<$ 0.06 &  $<$ 0.09 &  0.11 $\pm$ 0.03 &  0.25 $\pm$ 0.03 &  0.62 $\pm$ 0.11 &  1.26 $\pm$ 0.09 &  $<$ 0.16 \\		
& 0--3 & p7\tablenotemark{d} & \nodata & \nodata  &  $<$ 0.06 &  $<$ 0.09 &  0.10 $\pm$ 0.03 &  0.25 $\pm$ 0.03 &  0.58 $\pm$ 0.10 &  1.22 $\pm$ 0.10 &  $<$ 0.16 \\
& 4 && 16h20m29.19s & -25d25m47.91s		  &  $<$ 0.08 &  $<$ 0.10 &  $<$ 0.03 &  0.13 $\pm$ 0.03 &  0.96 $\pm$ 0.10 & 0.99 $\pm$ 0.11 &  $<$ 0.17 \\			
& 5 && 16h20m30.06s & -25d26m4.15s		  &  $<$ 0.07 &  $<$ 0.09 &  $<$ 0.04 &  0.15 $\pm$ 0.03 &  0.94 $\pm$ 0.11 & 0.99 $\pm$ 0.09 &  $<$ 0.19 \\			
& 6 && 16h20m28.56s & -25d26m18.76s		  &  $<$ 0.07 &  $<$ 0.10 &  $<$ 0.04 &  0.14 $\pm$ 0.03 &  0.90 $\pm$ 0.10 & 0.93 $\pm$ 0.09 &  $<$ 0.18 \\			
& 7 && 16h20m27.69s & -25d26m2.54s		  &  $<$ 0.07 &  $<$ 0.09 &  $<$ 0.04 &  0.15 $\pm$ 0.02 &  0.91 $\pm$ 0.10 & 0.96 $\pm$ 0.09 &  $<$ 0.21 \\			
& 4--7 & p9\tablenotemark{d} & \nodata & \nodata  &  $<$ 0.07 &  $<$ 0.09 &  $<$ 0.03 &  0.14 $\pm$ 0.03 &  0.93 $\pm$ 0.10 &  0.97 $\pm$ 0.10 &  $<$ 0.18 \\
& 8 && 16h20m38.61s & -25d27m55.01s		  &  $<$ 0.07 &  $<$ 0.10 &  $<$ 0.03 &  0.10 $\pm$ 0.03 &  0.95 $\pm$ 0.11 & 0.85 $\pm$ 0.10 &  $<$ 0.20 \\			
& 9 && 16h20m39.48s & -25d28m11.25s		  &  $<$ 0.07 &  $<$ 0.10 &  $<$ 0.04 &  0.11 $\pm$ 0.03 &  0.88 $\pm$ 0.11 & 0.82 $\pm$ 0.11 &  $<$ 0.18 \\			
& 10 && 16h20m37.98s & -25d28m25.86s		  &  $<$ 0.06 &  $<$ 0.11 &  $<$ 0.03 &  0.12 $\pm$ 0.02 &  0.91 $\pm$ 0.10 &  0.83 $\pm$ 0.10 &  $<$ 0.16 \\			
& 11 && 16h20m37.11s & -25d28m9.65s		  &  $<$ 0.06 &  $<$ 0.12 &  $<$ 0.03 &  0.10 $\pm$ 0.03 &  0.97 $\pm$ 0.10 &  0.93 $\pm$ 0.10 &  $<$ 0.19 \\			
& 8--11 & p10\tablenotemark{d} & \nodata & \nodata  &  $<$ 0.06 &  $<$ 0.11 &  $<$ 0.03 &  0.11 $\pm$ 0.02 &  0.93 $\pm$ 0.10 &  0.86 $\pm$ 0.10 &  $<$ 0.17 \\

NGC~1977 & 0 && 5h35m10.43s & -4d54m44.26s  &  $<$ 0.35 &  $<$ 0.32 &  2.50 $\pm$ 0.12 &  0.87 $\pm$ 0.23 &  $<$ 1.41 & 26.11 $\pm$ 1.10 &  $<$ 2.24 \\	  
& 1 && 5h35m11.49s & -4d54m56.29s	   &  $<$ 0.28 &  $<$ 0.42 &  2.50 $\pm$ 0.14 &  1.02 $\pm$ 0.20 &  $<$ 1.63 & 29.73 $\pm$ 1.37 &  $<$ 3.12 \\	  
& 2 && 5h35m10.49s & -4d55m16.20s	   &  $<$ 0.22 &  $<$ 0.36 &  1.33 $\pm$ 0.14 &  1.26 $\pm$ 0.17 &  $<$ 1.40 & 20.36 $\pm$ 0.98 &  $<$ 2.32 \\	  
& 3 && 5h35m9.42s & -4d55m4.19s		   &  $<$ 0.27 &  $<$ 0.39 &  1.47 $\pm$ 0.14 &  2.21 $\pm$ 0.14 &  $<$ 1.37 & 22.02 $\pm$ 1.10 &  $<$ 3.29 \\	  
& 0--3 & \cii\tablenotemark{e} & \nodata & \nodata  &  $<$ 0.21 &  $<$ 0.29 &  2.01 $\pm$ 0.10 &  1.24 $\pm$ 0.15 &  $<$ 1.20 &  24.53 $\pm$ 1.09 &  $<$ 2.27 \\ 
& 4 && 5h35m6.51s & -4d53m32.02s		   &  $<$ 0.30 &  $<$ 0.35 &  1.81 $\pm$ 0.16 &  1.14 $\pm$ 0.25 &  $<$ 1.63 & 21.72 $\pm$ 1.20 &  $<$ 3.50 \\	  
& 5 && 5h35m7.58s & -4d53m43.99s		   &  $<$ 0.23 &  $<$ 0.30 &  1.87 $\pm$ 0.18 &  $<$ 0.58 &  $<$ 1.83 &  23.33 $\pm$ 1.23 &  $<$ 3.68 \\		  
& 6 && 5h35m6.57s & -4d54m3.93s		   &  $<$ 0.49 &  $<$ 0.38 &  2.02 $\pm$ 0.14 &  1.99 $\pm$ 0.17 &  $<$ 1.20 & 22.79 $\pm$ 0.99 &  $<$ 3.06 \\	  
& 7 && 5h35m5.50s & -4d53m51.89s		   &  $<$ 0.28 &  $<$ 0.35 &  2.24 $\pm$ 0.13 &  2.71 $\pm$ 0.25 &  $<$ 1.51 & 24.75 $\pm$ 1.29 &  $<$ 2.49 \\	  
& 4--7 & \oi\tablenotemark{f} & \nodata & \nodata  &  $<$ 0.22 &  $<$ 0.27 &  2.03 $\pm$ 0.12 &  1.54 $\pm$ 0.21 &  $<$ 1.50 &  22.93 $\pm$ 1.06 &  $<$ 2.87 \\
& 8 && 5h35m5.11s & -4d53m52.84s		   &  $<$ 0.34 &  $<$ 0.26 &  2.12 $\pm$ 0.15 &  2.03 $\pm$ 0.20 &  $<$ 1.37 & 22.30 $\pm$ 1.04 &  $<$ 3.25 \\	  
& 9 && 5h35m6.18s & -4d54m4.92s		   &  $<$ 0.31 &  $<$ 0.29 &  1.79 $\pm$ 0.13 &  2.04 $\pm$ 0.20 &  $<$ 1.33 & 23.66 $\pm$ 0.95 &  $<$ 2.83 \\	  
& 10 && 5h35m5.17s & -4d54m24.80s	   &  $<$ 0.27 &  $<$ 0.26 &  1.44 $\pm$ 0.13 &  1.21 $\pm$ 0.18 &  $<$ 1.09 &  18.71 $\pm$ 0.77 &  $<$ 2.07 \\	  
& 11 && 5h35m4.11s & -4d54m12.79s	   &  $<$ 0.22 &  $<$ 0.30 &  1.56 $\pm$ 0.15 &  1.00 $\pm$ 0.16 &  $<$ 1.15 &  17.09 $\pm$ 0.67 &  $<$ 1.83 \\	  
& 8--11 & \oi\tablenotemark{g} & \nodata & \nodata  &  $<$ 0.20 &  $<$ 0.20 &  1.70 $\pm$ 0.12 &  1.56 $\pm$ 0.15 &  $<$ 0.99 &  20.19 $\pm$ 0.85 &  $<$ 2.25 \\
\tableline
\multicolumn{12}{l}{group b}\\
\tableline
G0.572-0.628 & 0 && 17h49m27.51s & -28d46m41.20s  &  4.70 $\pm$ 0.20 &  0.55 $\pm$ 0.18 &  $<$ 0.46 &  2.53 $\pm$ 0.36 &  459.85 $\pm$ 4.98 &  184.76 $\pm$ 4.18 &  3.96 $\pm$ 0.45 \\ 
& 1 && 17h49m27.90s & -28d46m36.01s &  4.90 $\pm$ 0.17 &  $<$ 0.57 &  0.59 $\pm$ 0.19 &  2.36 $\pm$ 0.33 &  477.74 $\pm$ 4.63 &  192.38 $\pm$ 4.12 &  4.13 $\pm$ 0.40 \\
& 2 && 17h49m27.51s & -28d46m11.13s &  4.11 $\pm$ 0.24 &  $<$ 0.56 &  $<$ 0.80 &  2.32 $\pm$ 0.37 &  368.34 $\pm$ 2.94 &  118.93 $\pm$ 2.45 &  5.60 $\pm$ 0.41 \\
& 3 && 17h49m27.90s & -28d46m5.99s &  3.71 $\pm$ 0.22 &  $<$ 0.69 &  $<$ 0.54 &  2.31 $\pm$ 0.35 &  352.23 $\pm$ 3.31 &  114.50 $\pm$ 2.38 &  6.05 $\pm$ 0.57 \\
& 4 && 17h49m27.51s & -28d45m41.12s &  2.96 $\pm$ 0.44 &  $<$ 0.82 &  $<$ 0.75 &  2.51 $\pm$ 0.45 &  242.27 $\pm$ 2.51 &  61.12 $\pm$ 1.25 &  6.66 $\pm$ 0.58 \\
& 5 && 17h49m27.90s & -28d45m35.99s &  2.99 $\pm$ 0.36 &  $<$ 0.94 &  $<$ 0.87 &  2.35 $\pm$ 0.52 &  243.49 $\pm$ 2.50 &  60.89 $\pm$ 1.34 &  5.92 $\pm$ 0.67 \\
& 6 && 17h49m27.51s & -28d45m11.11s &  3.21 $\pm$ 0.35 &  $<$ 1.10 &  $<$ 1.00 &  2.28 $\pm$ 0.39 &  266.13 $\pm$ 3.03 &  81.54 $\pm$ 1.67 &  6.01 $\pm$ 0.59 \\
& 7 && 17h49m27.90s & -28d45m5.98s &  3.85 $\pm$ 0.32 &  $<$ 0.72 &  $<$ 0.81 &  2.12 $\pm$ 0.56 &  291.83 $\pm$ 3.32 &  85.77 $\pm$ 1.72 &  6.80 $\pm$ 0.71 \\
G3.270-0.101 & 0 && 17h53m38.02s & -26d10m51.82s &  1.07 $\pm$ 0.12 &  0.61 $\pm$ 0.12 &  0.55 $\pm$ 0.15 &  4.18 $\pm$ 0.32 &  158.82 $\pm$ 1.67 &  41.32 $\pm$ 0.95 &  $<$ 1.09 \\
& 1 && 17h53m38.39s & -26d10m46.58s &  1.06 $\pm$ 0.11 &  0.61 $\pm$ 0.13 &  0.51 $\pm$ 0.13 &  4.30 $\pm$ 0.35 &  163.48 $\pm$ 1.69 &  41.87 $\pm$ 1.00 &  $<$ 0.99 \\
& 2 && 17h53m35.41s & -26d11m16.64s &  1.04 $\pm$ 0.23 &  $<$ 0.52 &  $<$ 0.60 &  4.18 $\pm$ 0.35 &  130.03 $\pm$ 1.22 &  27.73 $\pm$ 0.70 &  1.69 $\pm$ 0.37 \\
& 3 && 17h53m35.79s & -26d11m11.62s &  1.06 $\pm$ 0.24 &  $<$ 0.60 &  $<$ 0.69 &  4.25 $\pm$ 0.36 &  128.91 $\pm$ 1.43 &  30.56 $\pm$ 0.74 &  1.54 $\pm$ 0.35 \\
& 4 && 17h53m32.82s & -26d11m41.61s &  0.65 $\pm$ 0.11 &  0.41 $\pm$ 0.13 &  $<$ 0.25 &  3.85 $\pm$ 0.13 &  99.34 $\pm$ 1.15 &  17.45 $\pm$ 0.47 &  $<$ 0.96 \\
& 5 && 17h53m33.20s & -26d11m36.58s &  0.72 $\pm$ 0.11 &  $<$ 0.48 &  $<$ 0.28 &  4.00 $\pm$ 0.14 &  109.02 $\pm$ 1.24 &  19.05 $\pm$ 0.50 &  1.04 $\pm$ 0.29 \\
& 6 && 17h53m30.21s & -26d12m6.65s &  0.46 $\pm$ 0.11 &  $<$ 0.42 &  $<$ 0.24 &  3.71 $\pm$ 0.15 &  90.71 $\pm$ 1.06 &  21.71 $\pm$ 0.51 &  $<$ 0.84 \\
& 7 && 17h53m30.60s & -26d12m1.57s &  0.63 $\pm$ 0.10 &  $<$ 0.45 &  $<$ 0.36 &  3.67 $\pm$ 0.19 &  100.29 $\pm$ 1.07 &  21.37 $\pm$ 0.49 &  $<$ 0.83 \\
G4.412+0.118 & 0 && 17h55m20.89s & -25d4m42.40s  &  1.57 $\pm$ 0.19 &  0.74 $\pm$ 0.09 &  $<$ 0.29 &  2.16 $\pm$ 0.22 &  238.49 $\pm$ 1.71 &  41.65 $\pm$ 2.16 &  $<$ 3.73 \\
& 1 && 17h55m20.51s & -25d4m47.46s &  1.52 $\pm$ 0.18 &  0.59 $\pm$ 0.15 &  $<$ 0.43 &  2.26 $\pm$ 0.26 &  205.74 $\pm$ 1.48 &  38.45 $\pm$ 2.13 &  $<$ 3.80 \\
& 2 && 17h55m19.40s & -25d4m47.46s &  2.78 $\pm$ 0.39 &  $<$ 1.03 &  $<$ 0.78 &  2.01 $\pm$ 0.46 &  345.77 $\pm$ 2.83 &  48.99 $\pm$ 3.78 &  $<$ 6.37 \\
& 3 && 17h55m19.01s & -25d4m52.43s &  3.48 $\pm$ 0.80 &  $<$ 1.95 &  $<$ 1.87 &  $<$ 2.44 &  360.08 $\pm$ 2.78 &  45.60 $\pm$ 3.90 &  $<$ 6.93 \\
& 4 && 17h55m17.90s & -25d4m52.47s &  3.34 $\pm$ 1.05 &  $<$ 2.45 &  $<$ 2.35 &  $<$ 2.45 &  550.53 $\pm$ 4.98 &  88.72 $\pm$ 4.82 &  $<$ 9.88 \\       
& 5 && 17h55m17.51s & -25d4m57.41s &  3.09 $\pm$ 0.92 &  2.42 $\pm$ 0.76 &  $<$ 2.31 &  $<$ 2.07 &  493.67 $\pm$ 3.06 &  75.07 $\pm$ 4.20 &  $<$ 9.38 \\   
& 6 && 17h55m16.40s & -25d4m57.47s &  2.05 $\pm$ 0.33 &  1.49 $\pm$ 0.25 &  $<$ 0.87 &  2.43 $\pm$ 0.45 &  342.38 $\pm$ 2.28 &  64.08 $\pm$ 2.69 &  $<$ 4.50 \\
& 7 && 17h55m16.01s & -25d5m2.44s &  1.72 $\pm$ 0.40 &  1.17 $\pm$ 0.30 &  $<$ 0.99 &  2.41 $\pm$ 0.47 &  289.79 $\pm$ 2.50 &  53.37 $\pm$ 2.24 &  $<$ 3.95 \\
G8.137+0.228 & 0 && 18h3m0.82s & -21d48m13.30s  &  $<$ 13.70 &  18.85 $\pm$ 5.30 &  8.74 $\pm$ 2.57 &  $<$ 15.05 &  872.71 $\pm$ 20.11 &  138.58 $\pm$ 6.14 &  38.97 $\pm$ 9.20 \\ 
& 1 && 18h3m1.19s & -21d48m8.24s	&  24.27 $\pm$ 6.27 &  32.80 $\pm$ 9.62 &  $<$ 14.22 &  26.26 $\pm$ 6.85 &  1231.38 $\pm$ 24.88 &  175.79 $\pm$ 6.22 &  56.76 $\pm$ 10.39 \\
& 2 && 18h2m58.82s & -21d48m13.24s &  8.19 $\pm$ 1.21 &  6.80 $\pm$ 1.24 &  $<$ 2.91 &  4.08 $\pm$ 1.34 &  515.63 $\pm$ 6.04 &  100.90 $\pm$ 2.46 &  8.05 $\pm$ 2.50 \\
& 3 && 18h2m59.19s & -21d48m8.26s &  7.80 $\pm$ 1.03 &  5.48 $\pm$ 0.82 &  4.81 $\pm$ 1.03 &  $<$ 3.07 &  526.44 $\pm$ 7.10 &  116.54 $\pm$ 2.36 &  9.61 $\pm$ 3.00 \\
& 4 && 18h2m56.82s & -21d48m13.21s &  2.43 $\pm$ 0.32 &  2.81 $\pm$ 0.42 &  2.25 $\pm$ 0.32 &  3.02 $\pm$ 0.43 &  272.88 $\pm$ 3.44 &  80.75 $\pm$ 1.53 &  $<$ 2.46 \\
& 5 && 18h2m57.19s & -21d48m8.25s &  3.07 $\pm$ 0.36 &  2.98 $\pm$ 0.47 &  2.51 $\pm$ 0.41 &  2.82 $\pm$ 0.47 &  308.75 $\pm$ 4.06 &  84.89 $\pm$ 1.86 &  $<$ 2.96 \\
& 6 && 18h2m54.82s & -21d48m13.24s &  1.37 $\pm$ 0.15 &  2.08 $\pm$ 0.21 &  1.85 $\pm$ 0.18 &  2.66 $\pm$ 0.22 &  210.20 $\pm$ 2.32 &  75.96 $\pm$ 1.73 &  $<$ 1.70 \\
& 7 && 18h2m55.19s & -21d48m8.24s &  1.90 $\pm$ 0.13 &  2.13 $\pm$ 0.17 &  2.06 $\pm$ 0.20 &  2.77 $\pm$ 0.24 &  245.75 $\pm$ 2.41 &  82.61 $\pm$ 1.97 &  $<$ 1.83 \\
G32.797+0.192 & 0 && 18h50m31.32s & 0d1m49.21s  &  $<$ 17.83 &  $<$ 18.94 &  $<$ 20.14 &  $<$ 27.91 &  412.74 $\pm$ 16.08 & 182.67 $\pm$ 34.00 &  $<$ 99.17 \\	
& 1 & \tablenotemark{h} & 18h50m31.02s & 0d1m54.88s	&  $<$ 26.80 &  $<$ 28.07 &  $<$ 38.68 &  $<$ 43.62 &  422.10 $\pm$ 31.67 & 266.56 $\pm$ 52.70 &  $<$ 165.70 \\  
& 2 && 18h50m31.32s & 0d2m19.22s	&  0.58 $\pm$ 0.12 &  0.74 $\pm$ 0.20 &  1.84 $\pm$ 0.36 &  1.82 $\pm$ 0.50 &  40.78 $\pm$ 2.01 &  37.74 $\pm$ 2.94 &  $<$ 8.94 \\
& 3 && 18h50m31.02s & 0d2m24.85s	&  $<$ 0.44 &  $<$ 0.60 &  1.37 $\pm$ 0.32 &  2.59 $\pm$ 0.37 &  22.72 $\pm$ 1.56 &  27.36 $\pm$ 1.59 &  $<$ 7.08 \\
& 4 && 18h50m31.32s & 0d2m49.21s	&  $<$ 0.14 &  $<$ 0.21 &  0.32 $\pm$ 0.07 &  2.26 $\pm$ 0.10 &  6.86 $\pm$ 0.21 &  9.37 $\pm$ 0.46 &  $<$ 1.30 \\
& 5 && 18h50m31.02s & 0d2m54.88s	&  $<$ 0.16 &  $<$ 0.22 &  0.26 $\pm$ 0.07 &  2.29 $\pm$ 0.11 &  6.43 $\pm$ 0.23 &  9.08 $\pm$ 0.45 &  $<$ 1.24 \\
G48.930-0.286 & 0 && 19h22m14.40s & 14d3m3.14s  &  22.24 $\pm$ 0.56 &  12.75 $\pm$ 0.44 &  5.79 $\pm$ 0.40 &  $<$ 3.65 &  1065.36 $\pm$ 17.07 &  188.55 $\pm$ 8.69 &  81.83 $\pm$ 10.21 \\
& 1 && 19h22m14.78s & 14d3m7.97s	&  18.51 $\pm$ 0.39 &  13.07 $\pm$ 0.45 &  6.16 $\pm$ 0.51 &  $<$ 3.65 &  951.25 $\pm$ 9.24 &  179.55 $\pm$ 8.24 &  67.57 $\pm$ 9.26 \\
& 2 && 19h22m14.41s & 14d3m33.10s &  10.82 $\pm$ 0.27 &  8.59 $\pm$ 0.24 &  4.67 $\pm$ 0.45 &  $<$ 1.97 &  676.77 $\pm$ 11.10 &  153.24 $\pm$ 6.08 &  32.06 $\pm$ 6.32 \\
& 3 && 19h22m14.78s & 14d3m37.98s &  12.84 $\pm$ 0.41 &  10.54 $\pm$ 0.33 &  4.76 $\pm$ 0.55 &  $<$ 2.37 &  770.98 $\pm$ 17.05 &  159.63 $\pm$ 6.65 &  34.60 $\pm$ 7.09 \\
& 4 && 19h22m14.41s & 14d4m3.03s	&  15.18 $\pm$ 0.38 &  10.34 $\pm$ 0.41 &  7.96 $\pm$ 0.47 &  $<$ 3.50 &  877.91 $\pm$ 12.99 &  232.37 $\pm$ 10.74 &  43.24 $\pm$ 11.65 \\
& 5 && 19h22m14.77s & 14d4m8.01s	&  13.29 $\pm$ 0.45 &  9.41 $\pm$ 0.51 &  8.96 $\pm$ 0.62 &  $<$ 3.66 &  800.88 $\pm$ 13.64 &  229.16 $\pm$ 9.61 &  37.77 $\pm$ 9.20 \\
& 6 && 19h22m14.41s & 14d4m33.10s &  2.53 $\pm$ 0.11 &  2.60 $\pm$ 0.13 &  3.10 $\pm$ 0.12 &  1.88 $\pm$ 0.35 &  244.65 $\pm$ 1.94 &  86.71 $\pm$ 2.98 &  $<$ 6.83 \\
& 7 && 19h22m14.77s & 14d4m38.00s &  2.01 $\pm$ 0.08 &  1.90 $\pm$ 0.11 &  2.55 $\pm$ 0.13 &  2.04 $\pm$ 0.35 &  210.41 $\pm$ 2.61 &  75.33 $\pm$ 2.91 &  $<$ 5.93 \\
G79.293+1.296 & 0 && 20h28m8.83s & 40d52m17.95s  &  9.47 $\pm$ 0.55 &  3.73 $\pm$ 0.14 &  1.00 $\pm$ 0.24 &  $<$ 1.66 &  635.03 $\pm$ 4.78 &  64.08 $\pm$ 3.59 &  67.42 $\pm$ 6.41 \\
& 1 && 20h28m9.18s & 40d52m24.07s &  10.07 $\pm$ 0.72 &  4.82 $\pm$ 0.34 &  $<$ 1.46 &  $<$ 2.35 &  664.36 $\pm$ 5.72 &  68.40 $\pm$ 4.86 &  69.30 $\pm$ 8.35 \\    
& 2 && 20h28m8.13s & 40d51m48.03s &  12.03 $\pm$ 0.83 &  4.09 $\pm$ 0.50 &  5.58 $\pm$ 0.63 &  $<$ 2.81 &  758.90 $\pm$ 4.67 &  101.17 $\pm$ 6.25 &  89.07 $\pm$ 10.67 \\
& 3 && 20h28m8.47s & 40d51m54.08s &  12.13 $\pm$ 0.83 &  4.98 $\pm$ 0.51 &  6.55 $\pm$ 0.64 &  $<$ 2.83 &  751.05 $\pm$ 3.81 &  107.94 $\pm$ 5.72 &  85.70 $\pm$ 8.07 \\
& 4 && 20h28m7.83s & 40d51m18.03s &  11.06 $\pm$ 0.79 &  2.20 $\pm$ 0.47 &  $<$ 0.85 &  $<$ 2.28 &  659.60 $\pm$ 4.31 &  47.32 $\pm$ 3.71 &  85.03 $\pm$ 7.08 \\
& 5 && 20h28m8.17s & 40d51m24.08s &  10.76 $\pm$ 0.49 &  2.56 $\pm$ 0.26 &  $<$ 0.70 &  $<$ 2.19 &  620.02 $\pm$ 3.21 &  38.83 $\pm$ 3.59 &  85.64 $\pm$ 6.65 \\ 
& 6 && 20h28m7.13s & 40d50m48.01s &  4.11 $\pm$ 0.28 &  1.13 $\pm$ 0.21 &  1.49 $\pm$ 0.13 &  1.43 $\pm$ 0.44 &  372.17 $\pm$ 3.37 &  65.32 $\pm$ 3.22 &  38.66 $\pm$ 4.60 \\
& 7 && 20h28m7.48s & 40d50m54.09s &  5.44 $\pm$ 0.29 &  1.33 $\pm$ 0.35 &  1.30 $\pm$ 0.16 &  $<$ 1.59 &  454.20 $\pm$ 2.97 &  70.30 $\pm$ 3.50 &  48.79 $\pm$ 5.57 \\
G347.611+0.2 & 0 && 17h11m26.01s & -39d9m17.80s  &  $<$ 5.08 &  6.21 $\pm$ 1.51 &  $<$ 4.93 &  $<$ 5.85 &  499.40 $\pm$ 6.33 &  224.08 $\pm$ 6.05 &  $<$ 5.85 \\
& 1 && 17h11m26.45s & -39d9m12.87s &  $<$ 3.70 &  $<$ 3.98 &  $<$ 5.39 &  $<$ 6.05 &  506.73 $\pm$ 7.78 &  202.38 $\pm$ 5.84 &  $<$ 6.28 \\  
& 2 && 17h11m28.00s & -39d9m17.82s &  5.20 $\pm$ 1.03 &  7.36 $\pm$ 1.57 &  4.61 $\pm$ 1.16 &  $<$ 4.20 &  872.12 $\pm$ 9.23 &  224.30 $\pm$ 4.78 &  $<$ 8.03 \\
& 3 && 17h11m28.45s & -39d9m12.85s &  5.36 $\pm$ 1.18 &  8.80 $\pm$ 1.16 &  3.47 $\pm$ 1.13 &  $<$ 3.79 &  867.92 $\pm$ 8.88 &  209.42 $\pm$ 5.17 &  $<$ 6.53 \\
& 4 && 17h11m30.00s & -39d9m17.81s &  5.70 $\pm$ 0.94 &  8.40 $\pm$ 0.65 &  $<$ 2.36 &  4.46 $\pm$ 1.13 &  549.52 $\pm$ 7.12 &  73.04 $\pm$ 2.15 &  $<$ 3.64 \\ 
& 5 && 17h11m30.45s & -39d9m12.88s &  5.97 $\pm$ 0.90 &  10.06 $\pm$ 0.74 &  $<$ 2.08 &  4.76 $\pm$ 1.10 &  587.98 $\pm$ 7.50 &  73.57 $\pm$ 2.16 &  $<$ 3.82 \\
& 6 && 17h11m32.00s & -39d9m17.78s &  1.77 $\pm$ 0.44 &  3.19 $\pm$ 0.36 &  $<$ 1.38 &  3.20 $\pm$ 0.59 &  221.26 $\pm$ 3.65 &  34.39 $\pm$ 1.13 &  $<$ 1.70 \\
& 7 && 17h11m32.45s & -39d9m12.85s &  1.90 $\pm$ 0.45 &  3.34 $\pm$ 0.37 &  $<$ 1.35 &  2.90 $\pm$ 0.60 &  237.64 $\pm$ 4.34 &  34.87 $\pm$ 1.19 &  $<$ 1.73 \\
G351.467-0.462 & 0 & \tablenotemark{i} & 17h25m31.80s & -36d21m52.47s  &  8.69 $\pm$ 0.63 &  11.13 $\pm$ 0.58 &  4.55 $\pm$ 0.84 &  $<$ 3.99 &  828.56 $\pm$ 7.07 &  122.90 $\pm$ 8.10 &  $<$ 18.32 \\
& 1 && 17h25m32.21s & -36d21m47.33s &  8.56 $\pm$ 0.84 &  10.32 $\pm$ 1.08 &  4.24 $\pm$ 0.68 &  $<$ 3.29 &  739.57 $\pm$ 5.30 &  108.52 $\pm$ 6.07 &  $<$ 16.06 \\
& 2 && 17h25m33.79s & -36d21m52.52s &  4.54 $\pm$ 0.38 &  5.88 $\pm$ 0.45 &  3.90 $\pm$ 0.42 &  $<$ 1.68 &  488.36 $\pm$ 3.15 &  104.64 $\pm$ 4.40 &  $<$ 9.70 \\    
& 3 && 17h25m34.21s & -36d21m47.34s &  2.68 $\pm$ 0.21 &  3.82 $\pm$ 0.25 &  3.12 $\pm$ 0.22 &  1.42 $\pm$ 0.39 &  349.27 $\pm$ 2.15 &  90.73 $\pm$ 3.34 &  $<$ 6.66 \\
& 4 && 17h25m35.79s & -36d21m52.50s &  0.56 $\pm$ 0.08 &  1.92 $\pm$ 0.10 &  1.74 $\pm$ 0.14 &  1.38 $\pm$ 0.19 &  181.98 $\pm$ 1.58 &  59.49 $\pm$ 1.96 &  $<$ 3.03 \\
& 5 && 17h25m36.21s & -36d21m47.33s &  0.46 $\pm$ 0.06 &  1.38 $\pm$ 0.08 &  1.43 $\pm$ 0.09 &  1.29 $\pm$ 0.13 &  146.55 $\pm$ 0.87 &  52.12 $\pm$ 1.62 &  $<$ 2.37 \\
G359.429-0.090 & 0 && 17h44m36.16s & -29d28m0.16s  &  $<$ 5.43 &  7.06 $\pm$ 1.70 &  7.20 $\pm$ 2.37 &  $<$ 7.50 &  566.28 $\pm$ 7.49 &  280.89 $\pm$ 8.54 &  $<$ 11.35 \\
& 1 && 17h44m36.56s & -29d27m55.08s &  $<$ 5.13 &  5.25 $\pm$ 1.73 &  $<$ 6.15 &  $<$ 7.13 &  539.58 $\pm$ 6.59 &  294.23 $\pm$ 8.50 &  $<$ 10.59 \\
& 2 && 17h44m34.27s & -29d28m30.16s &  $<$ 3.10 &  5.25 $\pm$ 0.90 &  7.26 $\pm$ 1.00 &  6.53 $\pm$ 1.72 &  235.23 $\pm$ 5.03 &  265.74 $\pm$ 8.62 &  $<$ 6.61 \\
& 3 && 17h44m34.66s & -29d28m25.07s &  $<$ 1.15 &  3.33 $\pm$ 0.84 &  7.44 $\pm$ 1.25 &  8.30 $\pm$ 1.51 &  316.56 $\pm$ 6.39 &  306.54 $\pm$ 9.55 &  $<$ 7.96 \\
& 4\tablenotemark{j} && 17h44m32.37s & -29d29m0.14s &  $<$ 0.90 &  2.13 $\pm$ 0.38 &  3.53 $\pm$ 0.41 &  6.85 $\pm$ 0.47 &  112.41 $\pm$ 1.95 &  159.65 $\pm$ 5.87 &  $<$ 3.15 \\
& 5\tablenotemark{j} && 17h44m32.76s & -29d28m55.10s &  $<$ 0.92 &  2.15 $\pm$ 0.47 &  3.90 $\pm$ 0.43 &  6.90 $\pm$ 0.54 &  105.27 $\pm$ 1.91 &  161.79 $\pm$ 5.88 &  $<$ 3.47 \\
& 6\tablenotemark{j} && 17h44m30.47s & -29d29m30.16s &  $<$ 0.46 &  1.36 $\pm$ 0.15 &  2.93 $\pm$ 0.38 &  6.39 $\pm$ 0.44 &  106.38 $\pm$ 1.81 &  112.73 $\pm$ 4.32 &  $<$ 2.38 \\
& 7\tablenotemark{j} && 17h44m30.86s & -29d29m25.11s &  $<$ 0.39 &  1.37 $\pm$ 0.15 &  2.75 $\pm$ 0.38 &  6.50 $\pm$ 0.41 &  113.95 $\pm$ 2.11 &  118.74 $\pm$ 4.65 &  $<$ 2.58 \\
\enddata
\tablenotetext{a}{Identifiers in previous work.}
\tablenotetext{b}{Position in \citet{Okada03}.}
\tablenotetext{c}{Position in \citet{Simpson04}.}
\tablenotetext{d}{Position in \citet{Okada06}.}
\tablenotetext{e}{\cii\ peak position in \citet{YoungOwl02}.}
\tablenotetext{f}{\oi\ peak position in \citet{YoungOwl02}.}
\tablenotetext{g}{\oi\ second peak position in \citet{YoungOwl02}.}
\tablenotetext{h}{IRAS~18479-0005 in \citet{Peeters02}.}
\tablenotetext{i}{IRAS~17221-3619 in \citet{Peeters02}.}
\tablenotetext{j}{Affected by foreground contamination (see text).}
\end{deluxetable}

\clearpage
\pagestyle{plaintop}
\begin{deluxetable}{cccccccccc}
\tabletypesize{\scriptsize}
\rotate
\tablecaption{SH line intensities without extinction correction.\label{lineSH}}
\tablewidth{0pt}
\tablehead{
\colhead{Target} & \colhead{s\#} & \multicolumn{8}{c}{Line intensities [$10^{-8}$\,W\,m$^{-2}$\,sr$^{-1}$]}\\
\colhead{} & \colhead{} & \colhead{\suiv} & \colhead{H$_2$} & \colhead{\neii} & \colhead{\neiii} & \colhead{H$_2$} & \colhead{\piii} & \colhead{\feii} & \colhead{\suiii}\\
\colhead{} & \colhead{} & \colhead{10.5\um} & \colhead{12.3\um} & \colhead{12.8\um} & \colhead{15.5\um} & \colhead{17.0\um} & \colhead{17.9\um} & \colhead{17.9\um} & \colhead{18.7\um}
}
\startdata
\multicolumn{10}{l}{group a}\\
\tableline
S171 & 0 &  5.73 $\pm$ 0.12 &  2.76 $\pm$ 0.09 &  28.52 $\pm$ 0.64 &  13.35 $\pm$ 0.19 &  2.09 $\pm$ 0.08 &  0.32 $\pm$ 0.05 &  $<$ 0.10 &  27.46 $\pm$ 0.83 \\
& 1 &  4.94 $\pm$ 0.10 &  2.81 $\pm$ 0.07 &  28.23 $\pm$ 0.57 &  12.17 $\pm$ 0.17 &  2.08 $\pm$ 0.08 &  0.30 $\pm$ 0.04 &  $<$ 0.12 &  27.33 $\pm$ 0.78 \\
& 2 &  4.52 $\pm$ 0.08 &  2.48 $\pm$ 0.09 &  28.02 $\pm$ 0.58 &  11.73 $\pm$ 0.17 &  1.98 $\pm$ 0.08 &  0.25 $\pm$ 0.06 &  $<$ 0.11 &  26.66 $\pm$ 0.79 \\
& 3 &  4.90 $\pm$ 0.10 &  2.48 $\pm$ 0.07 &  28.36 $\pm$ 0.62 &  12.27 $\pm$ 0.18 &  2.07 $\pm$ 0.08 &  0.29 $\pm$ 0.05 &  $<$ 0.12 &  25.60 $\pm$ 0.77 \\
& 0--3 &  4.95 $\pm$ 0.10 &  2.56 $\pm$ 0.06 &  28.26 $\pm$ 0.60 &  12.24 $\pm$ 0.18 &  2.02 $\pm$ 0.07 &  0.29 $\pm$ 0.04 &  $<$ 0.09 &  26.75 $\pm$ 0.78 \\
& 4 &  4.26 $\pm$ 0.09 &  0.81 $\pm$ 0.08 &  27.49 $\pm$ 0.58 &  11.41 $\pm$ 0.16 &  0.94 $\pm$ 0.04 &  0.22 $\pm$ 0.07 &  $<$ 0.10 &  25.05 $\pm$ 0.75 \\
& 5 &  4.03 $\pm$ 0.10 &  0.57 $\pm$ 0.10 &  28.27 $\pm$ 0.61 &  11.12 $\pm$ 0.16 &  0.87 $\pm$ 0.05 &  0.27 $\pm$ 0.04 &  $<$ 0.08 &  24.90 $\pm$ 0.74 \\
& 6 &  3.91 $\pm$ 0.09 &  0.56 $\pm$ 0.09 &  28.29 $\pm$ 0.58 &  10.75 $\pm$ 0.15 &  0.73 $\pm$ 0.04 &  0.32 $\pm$ 0.04 &  $<$ 0.10 &  25.99 $\pm$ 0.79 \\
& 7 &  4.15 $\pm$ 0.11 &  1.16 $\pm$ 0.08 &  27.59 $\pm$ 0.58 &  11.06 $\pm$ 0.16 &  1.11 $\pm$ 0.05 &  0.24 $\pm$ 0.04 &  $<$ 0.08 &  25.96 $\pm$ 0.77 \\
& 4--7 &  4.09 $\pm$ 0.09 &  0.75 $\pm$ 0.08 &  27.93 $\pm$ 0.59 &  11.09 $\pm$ 0.16 &  0.90 $\pm$ 0.04 &  0.26 $\pm$ 0.04 &  $<$ 0.06 &  25.52 $\pm$ 0.77 \\
& 8 &  3.67 $\pm$ 0.08 &  0.78 $\pm$ 0.10 &  28.96 $\pm$ 0.64 &  10.48 $\pm$ 0.15 &  0.81 $\pm$ 0.05 &  0.33 $\pm$ 0.04 &  $<$ 0.06 &  25.99 $\pm$ 0.82 \\
& 9 &  3.55 $\pm$ 0.07 &  0.31 $\pm$ 0.08 &  29.77 $\pm$ 0.63 &  10.20 $\pm$ 0.15 &  0.39 $\pm$ 0.04 &  0.36 $\pm$ 0.05 &  $<$ 0.10 &  26.03 $\pm$ 0.80 \\
& 10 &  3.38 $\pm$ 0.06 &  0.54 $\pm$ 0.07 &  27.23 $\pm$ 0.57 &  9.67 $\pm$ 0.13 &  0.61 $\pm$ 0.04 &  0.26 $\pm$ 0.06 &  $<$ 0.13 &  24.36 $\pm$ 0.74 \\
& 11 &  3.21 $\pm$ 0.09 &  1.18 $\pm$ 0.08 &  26.93 $\pm$ 0.59 &  9.60 $\pm$ 0.14 &  1.13 $\pm$ 0.05 &  0.26 $\pm$ 0.06 &  $<$ 0.09 &  24.00 $\pm$ 0.73 \\
& 8--11 &  3.45 $\pm$ 0.07 &  0.63 $\pm$ 0.08 &  28.20 $\pm$ 0.61 &  9.95 $\pm$ 0.14 &  0.69 $\pm$ 0.04 &  0.31 $\pm$ 0.05 &  $<$ 0.08 &  25.06 $\pm$ 0.78 \\
& 12 &  0.73 $\pm$ 0.09 &  4.17 $\pm$ 0.11 &  38.16 $\pm$ 0.89 &  3.80 $\pm$ 0.06 &  3.34 $\pm$ 0.08 &  0.20 $\pm$ 0.06 &  $<$ 0.10 &  25.85 $\pm$ 0.74 \\
& 13 &  0.46 $\pm$ 0.09 &  5.41 $\pm$ 0.15 &  28.58 $\pm$ 0.81 &  3.03 $\pm$ 0.07 &  4.59 $\pm$ 0.10 &  $<$ 0.15 &  $<$ 0.10 &  16.92 $\pm$ 0.49 \\
& 14 &  0.67 $\pm$ 0.08 &  4.43 $\pm$ 0.14 &  30.94 $\pm$ 0.81 &  3.60 $\pm$ 0.07 &  4.59 $\pm$ 0.09 &  $<$ 0.17 &  $<$ 0.12 &  20.48 $\pm$ 0.59 \\
& 15 &  0.69 $\pm$ 0.10 &  7.64 $\pm$ 0.12 &  38.16 $\pm$ 0.85 &  3.56 $\pm$ 0.07 &  5.36 $\pm$ 0.10 &  $<$ 0.13 &  $<$ 0.17 &  24.93 $\pm$ 0.72 \\
& 12--15 &  0.63 $\pm$ 0.08 &  5.24 $\pm$ 0.12 &  34.03 $\pm$ 0.83 &  3.51 $\pm$ 0.06 &  4.46 $\pm$ 0.09 &  $<$ 0.12 &  $<$ 0.10 &  22.26 $\pm$ 0.64 \\
& 16 &  0.69 $\pm$ 0.08 &  1.64 $\pm$ 0.13 &  29.93 $\pm$ 0.84 &  3.45 $\pm$ 0.07 &  2.08 $\pm$ 0.07 &  $<$ 0.12 &  $<$ 0.12 &  18.84 $\pm$ 0.50 \\
& 17 &  0.56 $\pm$ 0.09 &  15.94 $\pm$ 0.18 &  20.88 $\pm$ 0.60 &  3.29 $\pm$ 0.07 &  11.61 $\pm$ 0.24 &  $<$ 0.12 &  $<$ 0.09 &  14.28 $\pm$ 0.40 \\
& 18 &  0.47 $\pm$ 0.07 &  11.19 $\pm$ 0.21 &  18.65 $\pm$ 0.56 &  2.91 $\pm$ 0.06 &  8.77 $\pm$ 0.16 &  $<$ 0.12 &  $<$ 0.11 &  12.22 $\pm$ 0.35 \\
& 19 &  0.56 $\pm$ 0.12 &  9.70 $\pm$ 0.21 &  20.45 $\pm$ 0.73 &  2.85 $\pm$ 0.07 &  9.44 $\pm$ 0.17 &  $<$ 0.13 &  $<$ 0.09 &  12.57 $\pm$ 0.35 \\
& 16--19 &  0.55 $\pm$ 0.08 &  10.37 $\pm$ 0.18 &  21.61 $\pm$ 0.67 &  3.03 $\pm$ 0.06 &  9.06 $\pm$ 0.18 &  $<$ 0.10 &  $<$ 0.08 &  13.98 $\pm$ 0.39 \\
& 20 &  0.53 $\pm$ 0.08 &  5.48 $\pm$ 0.12 &  21.36 $\pm$ 0.56 &  2.81 $\pm$ 0.05 &  4.55 $\pm$ 0.12 &  $<$ 0.10 &  $<$ 0.10 &  13.57 $\pm$ 0.36 \\
& 21 &  $<$ 0.33 &  18.01 $\pm$ 0.24 &  27.01 $\pm$ 0.86 &  2.12 $\pm$ 0.06 &  12.38 $\pm$ 0.24 &  $<$ 0.12 &  $<$ 0.11 &  11.95 $\pm$ 0.32 \\
& 22 &  $<$ 0.44 &  21.33 $\pm$ 0.26 &  20.48 $\pm$ 0.86 &  1.91 $\pm$ 0.09 &  14.49 $\pm$ 0.32 &  $<$ 0.19 &  $<$ 0.12 &  8.73 $\pm$ 0.24 \\
& 23 &  0.64 $\pm$ 0.10 &  5.53 $\pm$ 0.13 &  36.27 $\pm$ 0.89 &  2.82 $\pm$ 0.06 &  4.94 $\pm$ 0.13 &  $<$ 0.14 &  $<$ 0.15 &  18.67 $\pm$ 0.51 \\
& 20--23 &  0.45 $\pm$ 0.10 &  12.62 $\pm$ 0.17 &  27.08 $\pm$ 0.92 &  2.42 $\pm$ 0.06 &  9.08 $\pm$ 0.19 &  $<$ 0.12 &  $<$ 0.10 &  12.70 $\pm$ 0.40 \\
G333.6-0.2 & 0 &  8.04 $\pm$ 1.04 &  25.94 $\pm$ 1.70 &  1055.49 $\pm$ 26.58 &  71.48 $\pm$ 1.10 &  13.24 $\pm$ 1.23 &  $<$ 3.52 &  $<$ 2.08 &  444.63 $\pm$ 16.29 \\
& 1 &  9.79 $\pm$ 0.74 &  12.36 $\pm$ 1.03 &  1669.95 $\pm$ 38.64 &  60.01 $\pm$ 1.18 &  6.25 $\pm$ 0.95 &  12.21 $\pm$ 1.33 &  $<$ 2.28 &  1118.98 $\pm$ 35.00 \\
& 2 &  59.31 $\pm$ 1.92 &  7.52 $\pm$ 2.35 &  2826.80 $\pm$ 64.57 &  188.38 $\pm$ 2.44 &  $<$ 5.75 &  21.74 $\pm$ 2.61 &  $<$ 3.17 &  2093.52 $\pm$ 67.94 \\
& 3 &  12.61 $\pm$ 1.01 &  18.77 $\pm$ 2.56 &  2823.01 $\pm$ 68.30 &  98.99 $\pm$ 2.46 &  $<$ 6.16 &  $<$ 9.12 &  $<$ 6.11 &  1555.05 $\pm$ 50.77 \\
& 0--3 &  12.35 $\pm$ 0.98 &  18.81 $\pm$ 1.47 &  2092.03 $\pm$ 48.63 &  96.54 $\pm$ 1.85 &  6.75 $\pm$ 1.69 &  11.97 $\pm$ 2.24 &  $<$ 3.10 &  1303.94 $\pm$ 42.58 \\
& 4 &  30.16 $\pm$ 1.20 &  $<$ 3.65 &  1580.10 $\pm$ 37.41 &  130.65 $\pm$ 2.38 &  $<$ 3.70 &  11.33 $\pm$ 1.24 &  $<$ 1.69 &  1135.74 $\pm$ 33.57 \\
& 5 &  39.28 $\pm$ 1.70 &  5.86 $\pm$ 0.89 &  2907.65 $\pm$ 63.62 &  195.08 $\pm$ 3.17 &  $<$ 3.86 &  17.97 $\pm$ 1.76 &  $<$ 2.23 &  1802.95 $\pm$ 50.53 \\
& 6 &  13.26 $\pm$ 0.68 &  35.45 $\pm$ 1.77 &  1786.12 $\pm$ 50.57 &  121.92 $\pm$ 1.97 &  17.84 $\pm$ 1.18 &  $<$ 2.78 &  $<$ 1.72 &  790.99 $\pm$ 27.67 \\
& 7 &  105.31 $\pm$ 2.67 &  9.31 $\pm$ 1.92 &  4760.16 $\pm$ 117.06 &  399.57 $\pm$ 4.72 &  $<$ 7.30 &  22.24 $\pm$ 3.40 &  $<$ 3.76 &  3054.66 $\pm$ 95.79 \\
& 4--7 &  35.82 $\pm$ 1.35 &  16.23 $\pm$ 1.21 &  2631.04 $\pm$ 66.72 &  190.68 $\pm$ 3.13 &  7.45 $\pm$ 1.46 &  15.93 $\pm$ 1.68 &  $<$ 2.05 &  1653.28 $\pm$ 48.45 \\
& 8 &  6.84 $\pm$ 0.40 &  3.43 $\pm$ 0.25 &  537.09 $\pm$ 11.54 &  30.87 $\pm$ 0.45 &  3.64 $\pm$ 0.26 &  5.72 $\pm$ 0.34 &  $<$ 0.54 &  434.74 $\pm$ 13.02 \\
& 9 &  5.69 $\pm$ 0.53 &  2.98 $\pm$ 0.26 &  461.30 $\pm$ 9.79 &  26.41 $\pm$ 0.46 &  3.72 $\pm$ 0.24 &  4.90 $\pm$ 0.24 &  $<$ 0.62 &  353.25 $\pm$ 10.78 \\
& 10 &  11.75 $\pm$ 0.62 &  2.81 $\pm$ 0.60 &  565.84 $\pm$ 11.88 &  42.02 $\pm$ 0.75 &  2.40 $\pm$ 0.73 &  5.73 $\pm$ 0.80 &  $<$ 1.62 &  470.27 $\pm$ 13.86 \\
& 11 &  10.48 $\pm$ 0.37 &  2.45 $\pm$ 0.41 &  405.87 $\pm$ 8.96 &  37.07 $\pm$ 0.52 &  3.15 $\pm$ 0.30 &  4.59 $\pm$ 0.44 &  $<$ 0.69 &  336.09 $\pm$ 9.62 \\
& 8--11 &  8.09 $\pm$ 0.40 &  3.01 $\pm$ 0.28 &  487.21 $\pm$ 10.75 &  33.88 $\pm$ 0.42 &  3.47 $\pm$ 0.26 &  5.39 $\pm$ 0.25 &  $<$ 0.54 &  393.29 $\pm$ 11.28 \\
& 12 &  2.13 $\pm$ 0.36 &  1.21 $\pm$ 0.35 &  280.22 $\pm$ 6.62 &  12.04 $\pm$ 0.22 &  2.53 $\pm$ 0.14 &  2.36 $\pm$ 0.21 &  $<$ 0.36 &  194.61 $\pm$ 5.52 \\
& 13 &  3.06 $\pm$ 0.32 &  3.20 $\pm$ 0.36 &  516.57 $\pm$ 7.97 &  14.27 $\pm$ 0.25 &  3.65 $\pm$ 0.17 &  6.14 $\pm$ 0.17 &  $<$ 0.49 &  399.57 $\pm$ 9.50 \\
& 14 &  1.38 $\pm$ 0.30 &  2.59 $\pm$ 0.36 &  218.80 $\pm$ 5.07 &  8.38 $\pm$ 0.22 &  4.02 $\pm$ 0.11 &  2.10 $\pm$ 0.24 &  $<$ 0.36 &  150.74 $\pm$ 4.47 \\
& 15 &  1.33 $\pm$ 0.37 &  3.77 $\pm$ 0.27 &  331.56 $\pm$ 6.77 &  8.89 $\pm$ 0.23 &  4.06 $\pm$ 0.19 &  2.73 $\pm$ 0.23 &  $<$ 0.47 &  206.22 $\pm$ 5.72 \\
& 12--15 &  1.89 $\pm$ 0.30 &  2.73 $\pm$ 0.24 &  322.77 $\pm$ 6.07 &  10.45 $\pm$ 0.18 &  3.74 $\pm$ 0.11 &  2.48 $\pm$ 0.21 &  $<$ 0.32 &  208.92 $\pm$ 5.42 \\
\tableline
\multicolumn{10}{l}{group b}\\
\tableline
G351.467-0.462 & 0 &  22.09 $\pm$ 0.95 &  6.33 $\pm$ 1.25 &  1155.35 $\pm$ 22.55 &  84.50 $\pm$ 1.28 &  $<$ 2.68 &  4.67 $\pm$ 0.93 &  $<$ 1.85 &  769.05 $\pm$ 20.82 \\
& 1 &  21.18 $\pm$ 0.62 &  6.80 $\pm$ 0.76 &  1119.52 $\pm$ 20.42 &  77.32 $\pm$ 1.19 &  5.01 $\pm$ 0.83 &  4.38 $\pm$ 0.78 &  $<$ 1.56 &  743.31 $\pm$ 19.90 \\
& 2 &  1.86 $\pm$ 0.58 &  5.94 $\pm$ 0.93 &  429.93 $\pm$ 8.86 &  10.65 $\pm$ 0.47 &  3.78 $\pm$ 0.37 &  1.58 $\pm$ 0.30 &  $<$ 1.14 &  216.61 $\pm$ 6.09 \\
& 3 &  $<$ 1.59 &  4.91 $\pm$ 0.69 &  392.18 $\pm$ 7.89 &  9.10 $\pm$ 0.44 &  3.46 $\pm$ 0.47 &  $<$ 1.35 &  $<$ 1.49 &  199.55 $\pm$ 5.04 \\
& 4 &  $<$ 1.75 &  4.73 $\pm$ 0.79 &  126.47 $\pm$ 2.73 &  1.37 $\pm$ 0.39 &  2.96 $\pm$ 0.28 &  $<$ 0.95 &  $<$ 1.00 &  52.84 $\pm$ 1.48 \\
& 5 &  $<$ 2.02 &  4.28 $\pm$ 0.64 &  130.84 $\pm$ 2.18 &  $<$ 1.39 &  3.07 $\pm$ 0.40 &  $<$ 1.22 &  $<$ 0.89 &  54.97 $\pm$ 1.54 \\
\enddata
\end{deluxetable}

\clearpage

\begin{deluxetable}{cccc}
\tabletypesize{\scriptsize}
\tablecaption{Ionization potential of relevant elements.\label{IPtable}}
\tablewidth{0pt}
\tablehead{
\colhead{Elements} & \multicolumn{3}{c}{Ionization potential [eV]}\\
\colhead{} & \colhead{first} & \colhead{second} & \colhead{third}
}
\startdata
Fe & 7.87 & 16.16 & 30.65\\
Si & 8.15 & 16.35 & 33.49\\
S & 10.36 & 23.33 & 34.83\\
N & 14.53 & 29.60 & 47.45\\
Ar & 15.76 & 27.63 & 40.74\\
Ne & 21.56 & 40.96 & 63.45\\
\enddata
\end{deluxetable}

\begin{deluxetable}{cccc}
\tabletypesize{\scriptsize}
\tablecaption{Properties of relevant fine structure lines.\label{lineproptable}}
\tablewidth{0pt}
\tablehead{
\colhead{Ions} & \colhead{Wavelength} & \colhead{Configuration} & \colhead{Energy level of}\\
\colhead{} & \colhead{[\um]} &\colhead{} & \colhead{upper state [K]}
}
\startdata
\suiv & 10.51 & $^2$P$_{3/2}$-$^2$P$_{1/2}$ & 1369 \\
\neii & 12.81 & $^2$P$_{3/2}$-$^2$P$_{1/2}$ & 1123\\
\neiii & 15.55 & $^3$P$_1$-$^3$P$_2$ & 925 \\
       & 36.02 & $^3$P$_0$-$^3$P$_1$ & 1325 \\
\piii & 17.89 & $^2$P$_{3/2}$-$^2$P$_{1/2}$ & 804\\
\ariii & 21.84 & $^3$P$_0$-$^3$P$_1$ & 2259\\
\feiii & 22.93 & $^5$D$_3$-$^5$D$_4$ & 627 \\
\feii & 25.99 & $^6$D$_{7/2}$-$^6$D$_{9/2}$ & 554\\
      & 17.94 & $^4$F$_{7/2}$-$^4$F$_{9/2}$ & 3497\\
\suiii & 33.46 & $^3$P$_1$-$^3$P$_0$ & 430\\
       & 18.71 & $^3$P$_2$-$^3$P$_1$ & 1199\\
\siii & 34.81 & $^2$P$_{3/2}$-$^2$P$_{1/2}$ & 413\\
\enddata
\end{deluxetable}

\begin{deluxetable}{ccccccccc}
\tabletypesize{\scriptsize}
\tablecaption{Correlation coefficient and the t-test results of Figs.~\ref{SIIIH2SiII35} and \ref{SIIIH2FeII26}.\label{corr_coeff}}
\tablewidth{0pt}
\tablehead{
\colhead{Target} & \multicolumn{2}{c}{\siii\ 35\um} & \multicolumn{2}{c}{\siii\ 35\um} & \multicolumn{2}{c}{\feii\ 26\um} & \multicolumn{2}{c}{\feii\ 26\um}\\
\colhead{} & \multicolumn{2}{c}{vs. \suiii\ 33\um} & \multicolumn{2}{c}{vs. H$_2$ S(0) 28\um} & \multicolumn{2}{c}{vs. \suiii\ 33\um} & \multicolumn{2}{c}{vs. H$_2$ S(0) 28\um}\\
\colhead{} & \colhead{r\tablenotemark{a}} & \colhead{t [\%]\tablenotemark{b}} & \colhead{r\tablenotemark{a}} & \colhead{t[\%]\tablenotemark{b}} & \colhead{r\tablenotemark{a}} & \colhead{t[\%]\tablenotemark{b}} & \colhead{r\tablenotemark{a}} & \colhead{t[\%]\tablenotemark{b}}
}
\startdata
S171           & -0.60 & $<1$       & 0.61  & $<1$       & -0.79   & $<1$       & 0.71    & $<1$ \\
G333.6-0.2     & 0.80  & $<1$       & -0.22 & $>50$      & 0.69    & $2$--$5$   & -0.95   & $5$ \\
$\sigma$ Sco   & -0.68 & $1$--$2$   & 0.78  & $<1$       & -0.79   & $20$--$30$ & -0.92   & $5$--$10$ \\
NGC~1977       & \nodata & \nodata  & 0.09  & $>50$      & \nodata & \nodata    & 0.03    & $>50$ \\
G0.572-0.628   & 0.99  & $<1$       & 0.18  & $>50$      & \nodata & \nodata    & \nodata & \nodata \\
G3.270-0.101   & 0.95  & $<1$       & 0.64  & $5$--$10$  & \nodata & \nodata    & \nodata & \nodata \\
G4.412+0.118   & 0.91  & $<1$       & 0.39  & $>50$      & \nodata & \nodata    & \nodata & \nodata \\
G8.137+0.228   & 0.99  & $<1$       & 0.98  & $<1$       & 0.99    & $<1$       & 0.89    & $10$--$20$ \\
G32.797+0.192  & 0.99  & $<1$       & -0.61 & $30$--$40$ & 0.99    & $1$--$2$   & -0.67   & $30$--$40$ \\
G48.930-0.286  & 0.86  & $<1$       & \nodata & \nodata  & 0.72    & $2$--$5$   & \nodata & \nodata \\
G79.293+1.296  & 0.40  & $30$--$40$ & \nodata & \nodata  & 0.79    & $10$--$20$ & \nodata & \nodata \\
G347.611+0.2   & 0.71  & $2$--$5$   & 0.99  & $1$--$2$   & \nodata & \nodata    & \nodata & \nodata \\
G351.467-0.462 & 0.94  & $<1$       & 0.99  & $5$--$10$  & 0.97    & $<1$       & 0.99    & $5$--$10$ \\
G359.429-0.090 & 0.23  & $>50$      & \nodata & \nodata  & 0.99    & $5$--$10$  & \nodata & \nodata \\
\enddata
\tablecomments{Correlation coefficients are estimated only for targets where two lines are detected at more than two positions.}
\tablenotetext{a}{Correlation coefficient.}
\tablenotetext{b}{Rejection region of t-test.}
\end{deluxetable}

\clearpage

\begin{deluxetable}{ccc}
\tabletypesize{\scriptsize}
\tablecaption{Solar abundance and the gas-phase abundance in PDR models.\label{abundance_table}}
\tablewidth{0pt}
\tablehead{
\colhead{Element} & \colhead{Abundance} & \colhead{Ratio against solar}
}
\startdata
\multicolumn{3}{l}{Solar abundance by \citet{Asplund05}}\\
\tableline
Si & $3.26\times 10^{-5}$ & \nodata \\
Fe & $2.82\times 10^{-5}$ & \nodata \\
N & $6.03\times 10^{-5}$ & \nodata \\
O & $4.57\times 10^{-4}$ & \nodata \\
S & $1.38\times 10^{-5}$ & \nodata \\
\tableline
\multicolumn{3}{l}{Gas-phase abundance in PDR models by \citet{Kaufman06}}\\
\tableline
Si & $1.7\times 10^{-6}$ & 0.052 \\
Fe & $1.7\times 10^{-7}$ & 0.006 \\
O & $3.2\times 10^{-4}$ & 0.700 \\
\enddata
\end{deluxetable}

\clearpage

\begin{deluxetable}{cccccc}
\tabletypesize{\scriptsize}
\tablecaption{(Fe$^+$/Si$^+$)$_\mathrm{as}$, (Si$^{+}$/S$^{2+}$)$_\mathrm{as}$ and (Fe$^{2+}$/S$^{2+}$)$_\mathrm{as}$.\label{FeIII23SiII35SuIII33_table}}
\tablewidth{0pt}
\tablehead{
\colhead{Target} & \colhead{(Fe$^+$/Si$^{+}$)$_\mathrm{as}$} & \colhead{(Si$^+$/S$^{2+}$)$_\mathrm{as}$} & \colhead{(Fe$^{2+}$/S$^{2+}$)$_\mathrm{as}$}
}
\startdata
                S171  & $   0.065 $ $(+   0.044 -   0.040 $) & $   0.123 $ $(+   0.654 -   0.052 $) & $   0.013 $ $(+   0.008 -   0.009$) \\
          G333.6-0.2  & $   0.033 $ $(+   0.129 -   0.014 $) & $   0.218 $ $(+   0.326 -   0.103 $) & $   0.033 $ $(+   0.058 -   0.019$) \\
        $\sigma$ Sco  & $   0.228 $ $(+   0.153 -   0.084 $) & $   0.740 $ $(+   0.913 -   0.331 $) & \nodata \\
        G0.572-0.628  & $   0.007 $ $(+   0.003 -   0.002 $) & $   0.191 $ $(+   0.070 -   0.040 $) & $   0.003 $ $(+   0.001 -   0.001$) \\
        G3.270-0.101  & $   0.033 $ $(+   0.012 -   0.010 $) & $   0.129 $ $(+   0.031 -   0.030 $) & $   0.009 $ $(+   0.005 -   0.003$) \\
        G4.412+0.118  & $   0.032 $ $(+   0.013 -   0.009 $) & $   0.099 $ $(+   0.019 -   0.031 $) & $   0.008 $ $(+   0.008 -   0.003$) \\
        G8.137+0.228  & $   0.060 $ $(+   0.138 -   0.012 $) & $   0.158 $ $(+   0.083 -   0.072 $) & $   0.020 $ $(+   0.059 -   0.004$) \\
       G32.797+0.192  & $   0.091 $ $(+   0.069 -   0.043 $) & $   0.682 $ $(+   0.272 -   0.467 $) & $   0.039 $ $(+   0.018 -   0.012$) \\
       G48.930-0.286  & $   0.081 $ $(+   0.024 -   0.021 $) & $   0.149 $ $(+   0.087 -   0.045 $) & $   0.026 $ $(+   0.007 -   0.008$) \\
       G79.293+1.296  & $   0.061 $ $(+   0.109 -   0.034 $) & $   0.066 $ $(+   0.050 -   0.031 $) & $   0.011 $ $(+   0.007 -   0.007$) \\
        G347.611+0.2  & $   0.047 $ $(+   0.020 -   0.020 $) & $   0.113 $ $(+   0.168 -   0.041 $) & $   0.030 $ $(+   0.015 -   0.016$) \\
      G351.467-0.462  & $   0.079 $ $(+   0.042 -   0.016 $) & $   0.140 $ $(+   0.083 -   0.057 $) & $   0.025 $ $(+   0.013 -   0.006$) \\
      G359.429-0.090  & $   0.063 $ $(+   0.023 -   0.023 $) & $   0.390 $ $(+   0.346 -   0.098 $) & $   0.028 $ $(+   0.035 -   0.014$) \\
\enddata
\tablecomments{(Fe$^{+}$/Si$^{+}$)$_\mathrm{as}$ is derived from \feii\ 26\um/\siii\ 35\um, (Si$^{+}$/S$^{2+}$)$_\mathrm{as}$ is derived from \siii\ 35\um/\suiii\ 33\um, and (Fe$^{2+}$/S$^{2+}$)$_\mathrm{as}$ is derived from \feiii\ 23\um/\suiii\ 33\um.  We assume that all the emission comes from the ionized gas.  Errors indicate the range of extreme data points within each target.}
\end{deluxetable}

\begin{deluxetable}{cccccc}
\tabletypesize{\scriptsize}
\tablecaption{PDR properties from {\it ISO}/LWS observations.\label{PDRISO}}
\tablewidth{0pt}
\tablehead{
\colhead{Target} & \colhead{$T_d$\tablenotemark{a} [K]} & \colhead{$G_0$} & \colhead{$n$ [\cc]} & \colhead{$Z$\tablenotemark{b}} & \colhead{(Si$^+$)$_\mathrm{as}$}
}
\startdata
IRAS18479-0005 & 42--45 & 300--500 & 100--300 & 16 & 0.8--1.3\\
IRAS17221-3619 & 46--51 & 500--1000 & $\sim 1000$ & 5 & 1.0--2.0\\
\enddata
\tablenotetext{a}{Dust temperature.}
\tablenotetext{b}{Overlapping factor (see text and \citet{Okada03}).}
\end{deluxetable}

\begin{deluxetable}{cccccc}
\tabletypesize{\scriptsize}
\tablecaption{Fraction of \feii\ 26\um\ emission from the ionized gas.\label{ionizedFeII26}}
\tablewidth{0pt}
\tablehead{
\colhead{Target} & \colhead{s\#} &  \colhead{Fraction of \feii\ 26\um\ from the ionized gas}
}
\startdata
S171 & 13 & $<0.84$\\
& 19 & $<0.84$\\
& 20 & $<0.95$\\
& 21 & $<0.72$\\
& 22 & $<0.59$\\
G333.6-0.2 & 0 & $<0.45$\\
& 1 & $<0.80$\\
& 5 & $<0.72$\\
& 12 & $<0.57$\\
& 13 & $<0.90$\\
& 14 & $<0.99$\\
& 15 & $<0.96$\\
G351.467-0.462 & 0 & $<0.86$\\
& 1 & $<0.77$\\
& 2 & $<0.62$\\
\enddata
\tablecomments{Estimated by the upper limit of the \feii\ 18\um/26\um\ ratio.  Positions where \feii\ 26\um\ has not been detected or any constraint has not been made are omitted.}
\end{deluxetable}

\clearpage



\clearpage

\end{document}